\documentclass[letterpaper,final,5p,times,twocolumn,sort&compress]{elsarticle}
\usepackage{multirow}
\usepackage{amsmath,amssymb,bm,graphicx}
\usepackage{hyperref}
\usepackage{color}
\usepackage[usenames,dvipsnames]{xcolor}

\newcommand{\HS}{Hub\-bard-Stra\-to\-no\-vich}

\newcommand{\be}{\begin{equation}}
\newcommand{\ee}{\end{equation}}
\DeclareMathOperator{\tr}{tr}
\DeclareMathOperator{\Tr}{Tr}

\newcommand{\Uocc}{U_{\rm occ}}
\newcommand{\rhoocc}{\rho_{\rm occ}}
\newcommand{\Nocc}{N_{\rm occ}}
\newcommand{\Docc}{D_{\rm occ}}

\makeatletter
\def\ps@pprintTitle{%
  \let\@oddhead\@empty
  \let\@evenhead\@empty
  \def\@oddfoot{\reset@font\hfil\thepage\hfil}
  \let\@evenfoot\@oddfoot
}
\makeatother

\definecolor{DarkBlue}{RGB}{10,10,140}
\hypersetup{
    bookmarks=true,         
    colorlinks=true,        
    allcolors=DarkBlue
}

\begin{document}

\title{Reducing the complexity of finite-temperature auxiliary-field quantum Monte Carlo}


\author[1]{C.~N.~Gilbreth}\fnref{fn1}
\ead{christopher.gilbreth@gmail.com}
\author[2]{S.~Jensen}
\ead{scott.jensen@yale.edu}
\author[2]{Y.~Alhassid}
\ead{yoram.alhassid@yale.edu}
\address[1]{Institute for Nuclear Theory, Box 351550, University of Washington, Seattle, WA 98195}
\address[2]{Center for Theoretical Physics, Sloane Physics Laboratory, Yale University, New Haven, CT 06520}

\fntext[fn1]{Current address: Honeywell Quantum Solutions, 303 S Technology Ct., Broomfield, CO 80021, USA.}

\begin{abstract}
  The auxiliary-field quantum Monte Carlo (AFMC) method is a powerful and widely used technique for ground-state and finite-temperature simulations of quantum many-body systems. We introduce several algorithmic improvements for finite-temperature AFMC calculations of dilute fermionic systems that reduce the computational complexity of most parts of the algorithm.  This is principally achieved by reducing the number of single-particle states that contribute at each configuration of the auxiliary fields to a number that is of the order of the number of fermions.
 Our methods are applicable for both the canonical and grand-canonical ensembles. We demonstrate the reduced computational complexity of the methods for the homogeneous unitary Fermi gas. 
\end{abstract}

\maketitle

\section{Introduction}

Strongly interacting quantum many-body systems of fermions are ubiquitous in condensed-matter physics, nuclear physics, cold atom physics, and quantum chemistry.  Numerous theoretical approaches to their description exist, such as exact diagonalization, mean-field theory and its extensions, diagrammatic methods, density matrix renormalization group and tensor network algorithms, and quantum Monte Carlo (QMC) methods. QMC in particular is appealing in that, for certain classes of systems, it enables calculations that account for all quantum correlations of the system with a systematically controllable error.

The main advantage of QMC is that its computational effort scales gently (as a low power) in the number $N_s$ of single-particle states of the model space.
However, while offering a dramatic improvement to the exponential scaling of exact diagonalization methods, it is still computationally demanding and its practicality for larger values of $N_s$  depends sensitively on the particular power-law scaling in $N_s$.

Here we describe several algorithmic improvements for the finite-temperature auxiliary-field quantum Monte Carlo (AFMC) method. Principally, we introduce a novel method to identify, for a given field configuration, a subspace of mostly unoccupied single-particle states, and remove this subspace from the calculation. Since such states make only small contributions to observables, this can be done with a controlled error. Normally, the computational complexity of finite-temperature AFMC calculations scales as $O (N_s^3)$ when updating the auxiliary fields, and $O (N_s^3)$ or higher for the calculation of observables. Our method reduces this scaling in most parts of the algorithm by reducing the number of contributing single-particle states to a number $N_{\rm occ}$ of occupied states which depends weakly on  the original number of single-particle states $N_s$ and is usually of the order of the number of particles $N$. This enables otherwise impractical calculations, in particular in systems where $N_s \gg N$. We have applied this method to precision thermodynamic studies on the lattice of strongly interacting cold Fermi gases, including the pseudogap phenomenon~\cite{Jensen_review,Jensen2020_prl} and thermodynamic observables in the continuum limit~\cite{Jensen_contact}. The present paper is a detailed description of the method that was applied but only briefly discussed in Refs.~\cite{Jensen2020_prl,Jensen_review,Jensen_contact}.  Our method is applicable for both the canonical and grand-canonical ensembles.

The outline of this article is as follows. In Sec.~1, we briefly review the finite-temperature AFMC method for fermions, including particle-number projection and numerical stabilization. In Sec.~2, we discuss the method of reducing the single-particle model space based on the QR decomposition of the single-particle propagator, and show how to use this to efficiently compute one-body observables in the grand-canonical and canonical ensembles. In Sec.~3, we discuss an efficient method of updating the auxiliary fields. In Sec.~4, we consider two-body observables and describe a method to compute them efficiently in the canonical ensemble. In Sec.~5, we summarize the methods discussed here to improve the computational  complexity of the AFMC method. Finally in Sec.~6, we present our conclusion.

\subsection{AFMC method}

The finite-temperature AFMC method~\cite{Koonin1997_physrep,Alhassid2001_ijmpb,Alhassid2017} is based on the {\HS} transformation~\cite{Hubbard1959,Stratonovich1957}, in which the thermal propagator $e^{- \beta \hat{H}}$, where $\hat{H}$ is the Hamiltonian and $\beta = 1 / T$ is the inverse temperature, is expressed as a path integral of a one-body propagator $\hat{U} (\sigma)$ with respect to a weight $G(\sigma)$ over auxiliary fields $\sigma$:
\[ e^{- \beta \hat{H}} = \int D [\sigma] G (\sigma)  \hat{U} (\sigma) \;. \]
The auxiliary fields $\sigma$ are real or complex-valued quantities dependent on imaginary time $\tau$ and other indices. The imaginary time between $\tau=0$ and $\tau=\beta$  is usually discretized into $N_t+1$ equally spaced times $\tau_k = k \Delta \beta$  ($k=0,1,\ldots, N_t$) where $\Delta \beta = \beta / N_t$. The propagator $\hat U(\sigma)$ describes a system of non-interacting particles moving in one-body external auxiliary fields, and can be written as a product of propagators for each time step, $\hat{U} (\sigma) = \hat{U}_{N_t} \cdots \hat{U}_1$. In the following we will write $\hat{U} (\sigma) = \hat{U}$, omitting the explicit $\sigma$ dependence for simplicity.

The thermal expectation value of an observable $\hat{\mathcal O}$ is calculated from
\begin{equation} \label{thermExpect}
  \langle \hat{\mathcal{O}} \rangle = \frac{\Tr (e^{- \beta \hat{H}}
    \hat{\mathcal{O}})}{\Tr (e^{- \beta \hat{H}})} = \frac{\int D [\sigma] W (\sigma) \langle \hat{\mathcal{O}} \rangle_{\sigma} \Phi_{\sigma} }{\int D [\sigma] W (\sigma) \Phi_{\sigma}},
\end{equation}
where $W(\sigma) = G(\sigma) |\Tr \hat U|$ is a positive-definite weight function, $\langle \hat{\mathcal{O}} \rangle_{\sigma} = \Tr (\hat{U} \hat{\mathcal{O}}) / \Tr \hat{U}$ is the expectation value of $\hat{\mathcal O}$ in a given configuration $\sigma$ of the auxiliary fields, and $\Phi_{\sigma} = \Tr \hat{U} / | \Tr \hat{U} |$ is the Monte Carlo sign function. The traces can be grand-canonical, canonical, or involve projections onto other quantum numbers (e.g., angular momentum~\cite{Alhassid2007} and parity).  Because $\hat{U}$ is a one-body propagator, the traces can be computed using matrix algebra in the single-particle space of dimension $N_s$, yielding a scaling of $O (N_s^3)$ for the grand-canonical ensemble. For example, if the system is represented on a cubic 3D spatial lattice, $N_s \propto N_L^3$, where $N_L$ is the number of lattice points in each linear dimension, and the computational time scales as $O(N_L^9)$. Projections usually increase the computational complexity of calculating the trace. In the following, we will denote grand-canonical traces by $\Tr$, and canonical traces for $N$ particles by $\Tr_N$. Matrix traces will be denoted by $\tr$.

The auxiliary fields $\sigma$ are sampled stochastically according to $W(\sigma)$ and the expectation value of an observable $\hat{\mathcal O}$ is estimated from
\[ \langle \hat{\mathcal{O}} \rangle \approx \frac{\sum_i \langle
   \hat{\mathcal{O}} \rangle_{\sigma_i} \Phi_{\sigma_i}}{\sum_i \Phi_{\sigma_i}}\;,\]
where $i$ indexes the samples.  Various algorithms can be used (Metropolis-Hastings, Hybrid Monte Carlo, etc.)~\cite{Assaad2018_HMC,Duane1987_HMC} to sample the fields. For more details of the AFMC method, see Refs.~\cite{Koonin1997_physrep,Alhassid2001_ijmpb,Alhassid2017}.

In the grand-canonical ensemble with chemical potential $\mu$,
\begin{equation}
  \label{gctr} \Tr (\hat{U} e^{\beta \mu \hat{N}}) = \det (I + U e^{\beta \mu}),
\end{equation}
where $U$ is an $N_s\times N_s$ matrix representing the propagator $\hat{U}$ in the space of single-particle states. The thermal expectation of a one-body observable $\hat{\mathcal{O} }= \sum_{ij} \mathcal{O}_{ij} a_i^{\dag} a_j$ (here $a_i^\dag$ creates a particle in single-particle state $|i\rangle$ and $\mathcal{O}_{ij} = \langle i | \hat{\mathcal{O}}| j\rangle$) at the field configuration $\sigma$ is given by
\begin{equation}
  \label{onebdygctr}
  \langle \hat{\mathcal{O}} \rangle_{\sigma} =
  \frac{\Tr (e^{\beta \mu \hat{N}} \hat{U}  \hat{\mathcal{O}})}{\Tr (e^{\beta \mu \hat{N}}\hat{U})} =
  \tr (\mathcal{O} \rho),
\end{equation}
where $\rho$ is the one-body density matrix
\begin{equation}\label{rho}
\rho_{i j} = \langle a_j^{\dag} a_i \rangle_{\sigma} = \left[(I + e^{- \beta \mu} U^{- 1})^{- 1}\right]_{ij}\;.
\end{equation}

Computing \eqref{gctr} and \eqref{onebdygctr} each typically requires $O (N_s^3)$ operations. This scaling can be reduced for some systems by the method of pseudofermions, in which the determinant is computed by a stochastic sampling. This method is used extensively in lattice QCD calculations (see, for example in Ref.~\cite{Kennedy_qcd}). However, for nonrelativistic strongly interacting many-body systems, it does not always result in a lower overall computational effort~\cite{Assaad2018_HMC}, and computing the determinant directly is more common.

\subsection{Particle-number projection}\label{sec_numprojection}

The grand-canonical ensemble is effective for studying physical properties of bulk systems since the particle-number fluctuations relative to the average number of particles vanishes in the limit of large particle number. However, in finite-size systems such as atomic nuclei and metallic nanoparticles~\cite{Alhassid2013}, finite-size effects such as odd-even effects in particle number can be important and require the use of the canonical ensemble of fixed particle number.  It is therefore important to develop efficient quantum Monte Carlo methods in the framework of the the canonical ensemble.
The canonical ensemble is also important for studying physical properties of bulk systems which depend sensitively on particle number, such as pairing gaps. In particular, quantum Monte Carlo simulations in the canonical ensemble allow calculation of a finite-temperature gap from the staggering in energy with particle number, thereby providing direct information on pairing correlations without the need for analytic continuation~\cite{Gilbreth2013_pra,Jensen_review,Jensen2020_prl}.

Several methods for canonical-ensemble calculations at finite temperature have been used or proposed~\cite{Ormand1994_projection,Rombouts_canonical,Rehman_poly,Assaad2017_canonical,Drut2018_canonical}. We use an exact particle-number projection obtained by applying a discrete Fourier transform of grand-canonical traces for each field configuration. This is accomplished with the projection operator~\cite{Ormand1994_projection}
\begin{equation}
  \label{projop} \hat{P}_N = \frac{1}{N_s}  \sum_{m = 1}^{N_s} e^{i \varphi_m 
  (\hat{N} - N)} \;,
\end{equation}
where $\varphi_m = 2 \pi m / N_s$ for $m=1,\ldots, N_s$. Inserting $\hat{P}_N$ into the grand-canonical trace, we obtain the canonical partition function $Z_N = Z_N(\sigma)$ for a given field configuration
\begin{equation}
  \label{Zfourier}
 Z_N= {\Tr}_N \; \hat{U} = \Tr (\hat{P}_N  \hat{U}) =
  \frac{e^{- \beta \mu_C N}}{N_s} \sum_{m = 1}^{N_s} e^{- i \varphi_m N} \eta_m \;,
\end{equation}
where $\eta_m \equiv \Tr (e^{i \varphi_m \hat{N}} e^{\beta \mu_C \hat{N}} \hat{U}) = \det (I + e^{i \varphi_m} e^{\beta \mu_C} U)$.

Eq.~\eqref{Zfourier} contains an additional parameter $\mu_C = \mu_C(\sigma)$, which is a chemical potential for the non-interacting system described by $\hat U$~\cite{Ormand1994_projection}. In exact arithmetic the formula {\eqref{Zfourier}} holds for any value of $\mu_C$. However, it can be numerically unstable if $\mu_C$ is such that the average particle number in the grand-canonical ensemble, $\Tr(e^{\beta \mu_C \hat{N}} \hat{U} \hat{N})/\Tr(e^{\beta \mu_C \hat{N}} \hat{U})$, is very different than $N$. In this case the desired $N$-particle trace will be a very small contribution to the expansion of the grand-canonical trace $\Tr [e^{\beta \mu_C \hat N} \hat U(\sigma)]$ in powers of $e^{\beta\mu_C}$, and will be difficult to extract numerically.

To avoid this, one can determine $\mu_c$ at each field configuration so that the grand-canonical expectation of $\hat N$ is $N$, i.e.,
\begin{equation} \label{eq:muC}
\frac{\Tr (e^{\beta \mu_C \hat{N}} \hat{U} \hat{N})}{\Tr (e^{\beta \mu_C \hat{N}} \hat{U})} = N \,.
\end{equation}
This equation can be solved numerically with negligible computational overhead using a single-particle basis in which $\hat U(\sigma)$ is diagonal. In practice, $\mu_C$ need not be exact; a broad range of values tends to stabilize the Fourier sum.

Note $\mu_C$ should not be confused with a fixed chemical potential $\mu$ for the fully interacting system that would be introduced in Eq.~\eqref{thermExpect} to sample in the grand-canonical ensemble. Such a chemical potential is not needed to sample in the canonical ensemble. Instead, determining $\mu_c$ as described for each field configuration and using Eq.~\eqref{Zfourier} allows direct and exact sampling of the fields in the canonical ensemble. 

The canonical expectation value of an observable $\hat{\mathcal O}$ for $N$ particles in a given configuration $\sigma$ of the auxiliary fields can be computed from
\begin{equation}
  \label{xproj}  \langle \hat{\mathcal O} \rangle_{\sigma} = \frac{\Tr (\hat{P}_N \hat{U} \hat{\mathcal O})}{\Tr (\hat{P}_N  \hat{U})} = \frac{e^{- \beta \mu_C N}}{Z_N  N_s} \sum_{m = 1}^{N_s} e^{- i \varphi_m N}  \langle \hat{\mathcal O}\rangle_\sigma^{(m)}
  \eta_m \,,
\end{equation}
where $\langle \hat{\mathcal O} \rangle_\sigma^{(m)} = \Tr (e^{\beta \mu_C \hat{N}} e^{i \varphi_m \hat{N}} \hat{U} \hat{\mathcal O}) / \Tr (\hat{U} e^{\beta \mu_C \hat{N}} e^{i \varphi_m \hat{N}})$ and $Z_N$ is given by \eqref{Zfourier}.

In the case of one-body density operators, $\hat{\mathcal O} = a^{\dagger}_i a_j$ and 
\be\label{gamma-ij-m}
\langle a^{\dagger}_i a_j \rangle_\sigma^{(m)} = \left[(I + e^{- \beta \mu_C - i \varphi_m} U^{- 1} )^{- 1}\right]_{j i} \;.
\ee
  The thermal expectation value of two-body observables can be computed using Wick's theorem for the grand-canonical traces.

The calculation of projected quantities in Eqs.~\eqref{Zfourier} and \eqref{xproj}
can be accomplished in $O(N_s^3)$ operations by diagonalizing the matrix $U$~\cite{Gilbreth2015_CPC,Rombouts_canonical}. Given the eigenvalue decomposition $U= P \Lambda P^{-1}$, where $\Lambda_{i j} = \delta_{i j} \lambda_i$, we obtain
\begin{align}
    \eta_m & =  \prod_{k = 1}^{N_s} \left(1 + e^{\beta \mu_C}  e^{i \varphi_m} \lambda_k\right)\;, \label{etam} \\
    \rho_{\alpha} & =  \frac{e^{- \beta \mu_C N}}{Z_N N_s}
    \sum_{m = 1}^{N_s} e^{-i\varphi_m N} \frac{\lambda_{\alpha} e^{\beta \mu_C} e^{i \varphi_m}}{1 + \lambda_{\alpha} e^{\beta \mu_C} e^{i \varphi_m}} \eta_m, \label{rhoalphafourier}\\
  \rho_{i j} & =  \sum_{\alpha = 1}^{N_s} P_{i \alpha} \rho_{\alpha} P^{-1}_{\alpha j} .  \label{rhoijfourier}
\end{align}

The $O (N_s^3)$ complexity of the diagonalization or calculation of the determinant can still be problematic for larger values of $N_s$. In Sec.~\ref{sec_trunc} we discuss how to effectively reduce the dimension of the matrix $U$ in order to improve the scaling.

\subsection{Numerical stabilization}

AFMC simulations at large imaginary times involve long chains of matrix multiplications to compute $U = U_{N_{\tau}} \cdots U_1$, which can give rise to numerical instabilities when the matrices $U$ become ill-conditioned~\cite{Koonin1997_physrep,LohJr1992,Bai2011_laa}. To address this, one can accumulate a decomposition of $U$ which separates out the important scales. One effective choice is the QDR decomposition $U = Q D R$, where $Q$ is unitary, $D$ is diagonal with positive entries, and $R$ is unit right-triangular, obtained from a QR decomposition. Although in principle a pivoted QR decomposition more accurately separates the scales, in practice the QR decomposition without pivoting is highly accurate.
A QR decomposition need not be done for every imaginary time $\tau_k$, but can be done periodically, e.g., every $K$ time slices (for example $K = 16$). The numerical stabilization then requires $O ([N_t / K] N_s^3)$ operations for each update of all auxiliary fields.

To diagonalize $U$, one cannot multiply out the factors $Q D R$ in this order without destroying the information contained in the smaller scales of $D$.  Instead, one can diagonalize the matrix $S = D R Q$ obtained from $U$ by a similarity transformation, and from this calculate number-projected observables. This procedure reduces the number of operations requires for numerically stabilized canonical-ensemble calculations from $O(N_s^4)$ (without diagonalization) to $O (N_s^3)$~\cite{Gilbreth2015_CPC}.

\subsection{The unitary Fermi gas}\label{subsec_ufg}

 Our algorithm developments will be illustrated using a lattice model of the unitary Fermi gas (UFG), a strongly-interacting quantum many-body system with connections to diverse areas of physics, and which has been the subject of many theoretical and experimental investigations (see, e.g., Refs.~\cite{Zwerger2012,Randeria2014_review,Jensen_review} for recent reviews). The UFG is the thermodynamic limit of a three-dimensional system of spin-1/2 fermions interacting with an attractive contact interaction whose $s$-wave scattering length is infinite. It exhibits a superfluid phase transition at a critical temperature of $T_c \approx 0.17\, T_F$, where $T_F$ is the Fermi temperature of the gas. The UFG is the midpoint of a continuous crossover between a Bardeen-Cooper-Schrieffer (BCS) pairing regime and Bose-Einstein-Condensate (BEC) regime as the inverse scattering length is varied~\cite{Randeria2014_review}.

The UFG can be modeled by considering a finite number $N$ of spin-1/2 particles interacting in a fixed volume of space discretized into a cubic lattice with $N_L$ points in each dimension (for a total of $N_L^3$ lattice points), and taking the continuum limit $N_L \rightarrow \infty$ and thermodynamic limit $N \rightarrow \infty$. The lattice Hamiltonian of the UFG is given by 
\begin{equation} \label{ham}
\hat{H}=\sum_{\mathbf{k},s_z}\epsilon _{\mathbf{k}}\hat{a}^{\dagger }_{\mathbf{k},s_Z}\hat{a}_{\mathbf{k},s_z} + g\sum_{\mathbf{x}}\hat{n}_{\mathbf{x},\uparrow}\hat{n}_{\mathbf{x},\downarrow} \;,
\end{equation}
where $\hat{a}^{\dagger }_{\mathbf{k},s_z}$ and $\hat{a}_{\mathbf{k},s_Z}$ create and annihilate, respectively, a fermion with wavevector ${\bf k}$, spin projection $s_z= \uparrow, \downarrow$, and single-particle energy $\epsilon _{\mathbf{k}} = {\hbar^2 k}^{2} /2m$. The lattice site occupation number operator is  $\hat{n}_{\mathbf{x},s}=\hat{\psi}^{\dagger}_{\mathbf{x},s}\hat{\psi}_{\mathbf{x},s}$, where $\hat{\psi}^{\dagger}_{\mathbf{x},s}, \hat{\psi}_{\mathbf{x},s}$ obey fermionic anticommutation relations $\{ \hat{\psi}^{\dagger}_{\mathbf{x},s_Z},\hat{\psi}_{\mathbf{x}',s_z'}\}= \delta_{\mathbf{x},\mathbf{x}'}\delta_{s_z,s_z'}$. The negative coupling constant $g$ is tuned to the limit of infinite scattering length for each lattice size by solving the Lippman-Schwinger equation. See Ref.~\cite{Jensen2020_prl} for further details.

The calculations reported here are performed for lattice systems with varying lattice sizes and fixed particle number $N = N_\uparrow + N_\downarrow = 66$, similar parameters to those of Ref.~\cite{Jensen_contact}. Of particular interest is the temperature regime is just above and below the critical temperature for superfluidity, $0.05\, T_F \lesssim T \lesssim 0.25 \,T_F$. 

\section{Model space truncation}\label{sec_trunc}

In this section we introduce a novel method to reduce the dimension of the one-body propagator given the decomposition $U = Q D R$ from $N_s$ to $N_{\rm occ}$, where $N_{\rm occ}$ is of the order of number of particles $N$. This reduces the computational complexity of both grand-canonical and canonical calculations to below $O (N_s^3)$ for all parts of the algorithm except the QR decomposition used for stabilization, and greatly speeds up calculations.

\subsection{Omitting unoccupied states}

In the decomposition $U = Q D R$ for a given field configuration $\sigma$, the entries of the diagonal matrix $D$ correspond roughly to the eigenvalues of $U$. Small eigenvalues of $U$ correspond to mostly unoccupied single-particle states and contribute to observables proportionally to their magnitude. When their total contribution is below a desired accuracy threshold, these states can be omitted from the model space, thereby reducing the dimension of the single-particle space in which traces are calculated.

In the following we will not assume the matrices $Q$ and $R$ to be unitary or right-triangular, respectively, but instead treat them as general complex matrices. This is necessary as the $\mathrm{QR}$ decomposition is not performed at every imaginary time step.

\begin{figure}[t]
  \begin{center}
  \includegraphics[width=\columnwidth]{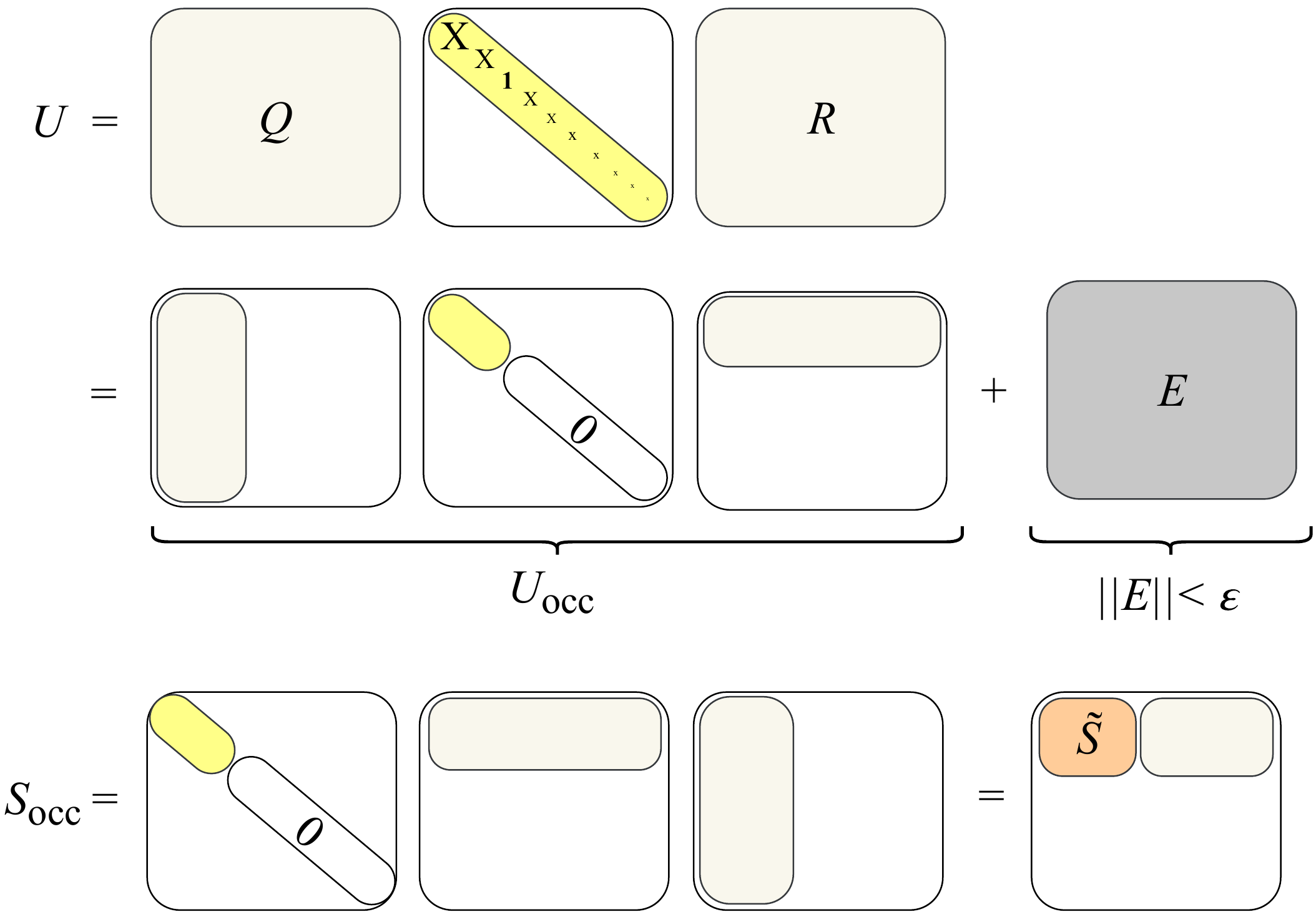}
  \caption{Reduction of the model space using the elements of the diagonal $D$ matrix (after it is scaled by $e^{\beta \mu}$ or $e^{\beta \mu_D}$). $U$ is separated into a matrix $\Uocc$ of rank $\Nocc$ plus a perturbation $E$, where $\Nocc$ is determined by the condition $||E|| < \varepsilon$. To compute observables the matrix $\tilde{S}$ is computed, which is the upper left $\Nocc \times \Nocc$ block of the matrix $S_{\rm occ} = D_{\rm occ} R Q$, where $D_{\rm occ}$ contains only the elements of $D$ corresponding to occupied states. \label{fig_truncation}}
  \end{center}
\end{figure}

The process of reducing the dimension of the single-particle space is illustrated in Fig.~\ref{fig_truncation}. Given the decomposition $Q D R$ of $U$ or a matrix similar to $U$, we first determine a reference scale for the Fermi energy at the given field configuration. In the grand-canonical ensemble, this is provided by the chemical potential $\mu$. In the canonical ensemble, we determine an approximate chemical potential $\mu_D$ from  the diagonal elements of $D$ by numerically solving for $\mu_D$ in the equation
\[ \sum_k \frac{d_k e^{\beta \mu_D}}{1 + d_k e^{\beta \mu_D}} = N, \]
where $D_{i j} = d_i \delta_{i j}$. 
In the following we will absorb the scale factor $e^{\beta \mu_D}$ or $e^{\beta \mu}$ (the latter for the case of the grand-canonical ensemble) into the elements of $D$ (and therefore $U$), so that the Fermi energy is located at $d_k \approx 1$.

Given this rescaling of $D$, we separate out a certain number of the smallest diagonal values of $D$ to reduce the rank of $U$. Sorting the entries of $D$ in an appropriate order (see below), we write  $D = D_{\rm occ} + D_{\varepsilon}$, where $D_{\rm occ} = \mathrm{diag} (d_1, \ldots, d_{N_{\mathrm{occ}}}, 0, 0, \ldots)$ and $D_{\varepsilon} = \mathrm{diag} (0, 0, \ldots, 0, d_{N_{\mathrm{occ}} + 1}, \ldots, d_{N_s})$, and $N_{\mathrm{occ}}$ is the number of significantly occupied states.  We then have
\begin{equation} \label{truncation}
  U = Q D_{\rm occ} R + Q D_{\varepsilon} R = U_{\rm occ} + E \;,
\end{equation}
where $U_{\rm occ} = Q D_{\rm occ} R$ is the ``occupied'' part of $U$
and $E = Q D_{\varepsilon} R$ is a small perturbation. Subsequent to determining the truncation in Eq.~\eqref{truncation}, all observables are calculated with $\Uocc$ rather than $U$, taking advantage of the reduced rank of $\Uocc$.

To determine $N_{\mathrm{occ}}$, we choose it to be a small integer such that $\| E \| < \varepsilon$, where $\varepsilon$ is a given small parameter and $\| \cdot \|$ is the Frobenius norm $\| E \| = \left( \sum_{i j} |E_{i j}|^2 \right)^{1/2}$.  This can be done with $O (N_s^2)$ operations using the upper bound
\begin{equation}
  \label{enrm} \| E \| \leq \| Q \| \, \| D_\varepsilon R \| = \| Q \| \left (\sum_{j=1}^{N_s} \sum_{k=N_{\rm occ}+1}^{N_s} d_k |R_{k j}|^2 \right)^{1/2} < \varepsilon \;.
\end{equation}
For each field configuration we choose $\Nocc$ to be the smallest integer which satisfies the rightmost inequality in Eq.~\eqref{enrm}. According to \eqref{enrm}, the most natural sorting order of the $d_k$ values is by the contributions $d_k \sum_{i = 1}^{N_s} | R_{k i} |^2$ to $\| E \|$ rather than the $d_k$ values themselves.

\begin{figure}[t!]
 \begin{center}
  \includegraphics[width=\columnwidth]{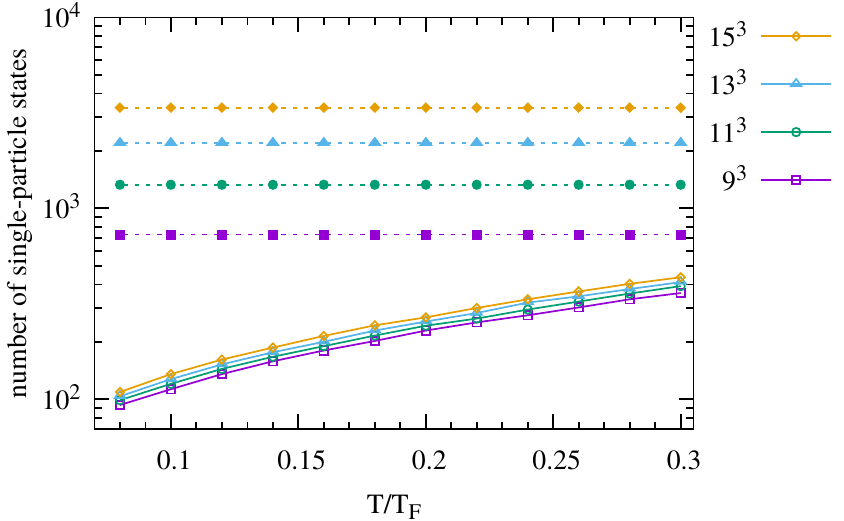}
  \caption{Dimensions $N_s = N_L^3$ of the complete single-particle model space (solid symbols connected by dashed lines) and $N_{\rm occ}$ of the reduced single-particle model space (open symbols connected by solid lines) for the unitary Fermi gas a function of $T/T_F$, where $T_F$ is the Fermi temperature, for multiple lattices. The simulations were carried out for the spin-balanced case with $N = 66$ particles (number projections were performed onto $N_{\uparrow}=N_{\downarrow}=33$ particles) and temperatures in a physically relevant region around the experimental superfluid critical temperature $T_c \approx 0.17  \,T_F$. To determine $N_{\rm occ}$ for the model space reduction we used $\varepsilon=10^{-6}$. The number of states shown is for a single species, i.e., spin-up ($\uparrow$) or spin-down ($\downarrow$). $N_{\rm occ}$ depends primarily on the number of particles and temperature, and only weakly on the lattice size.\label{fig_nred}}
  \end{center}
\end{figure}

The number $N_{\mathrm{occ}}$ of significantly occupied states will depend on the number of particles $N$ and the temperature $T$. For a given $N$ and $T$, $N_{\mathrm{occ}}$ will increase gently with the total number $N_s$ of single-particle states once $N_s$ is sufficiently large to capture the relevant physics. Fig.~\ref{fig_nred} shows on a logarithmic scale $N_s$ and $N_{\rm occ}$ vs.~$T/T_F$ (where $T_F$ is the Fermi temperature) for the spin-balanced unitary Fermi gas with $N = 66$ particles (i.e., $N_\uparrow=N_\downarrow=33$). Temperatures are shown in a physically interesting regime around the experimental superfluid critical temperature $T/T_F \approx 0.17$.  In general, $N_{\mathrm{occ}}$ is much smaller than $N_s$ in this regime, and is almost independent of the lattice size (for fixed particle number). As a result, the benefit of the truncation becomes greater for larger lattices.

To calculate observables we permute the factors $Q, D_{\rm occ}, R$ to obtain
\begin{equation} \label{Socc}
  S_{\rm occ} \equiv D_{\rm occ} R Q = \left(\begin{array}{ccccc}
  \cline{1-3}
     \multicolumn{1}{|c}{X} & X & \multicolumn{1}{c|}{X} & X & X\\
     \multicolumn{1}{|c}{\scriptstyle{X}} & \scriptstyle{X} & \multicolumn{1}{c|}{\scriptstyle{X}} & \scriptstyle{X} & \scriptstyle{X}\\
     \multicolumn{1}{|c}{\scriptscriptstyle{X}} & \scriptscriptstyle{X} & \multicolumn{1}{c|}{\scriptscriptstyle{X}} & \scriptscriptstyle{X} & \scriptscriptstyle{X}\\
     \cline{1-3}
     0 & 0 & 0 & 0 & 0\\
     0 & 0 & 0 & 0 & 0
\end{array}\right)_{N_s \times N_s},
\end{equation}
which is a similarity transformation $S_{\rm occ} = Q^{- 1} U_{\rm occ} Q$. The boxed block in Eq.~\eqref{Socc} is an $N_{\mathrm{occ}} \times N_{\mathrm{occ}}$ matrix which we denote by $\tilde{S}$. Computing the matrix elements of $\tilde{S}$ requires $O (N_s N_{\mathrm{occ}}^2)$ operations. 

\subsection{Grand-canonical-ensemble observables} \label{gcobs}

The grand-canonical partition function in \eqref{gctr} can then be approximated by $Z_{\rm occ} = \det (I + U_{\rm occ})$ (note that the factor $e^{\beta\mu}$ is absorbed into $D$ or $U$), and computed from $\tilde S$ by
\begin{equation} \label{Zocc}
  Z_{\rm occ} =  \det (I + \tilde{S}) \;,
\end{equation}
which requires $O (N_{\mathrm{occ}}^3)$ operations to compute. To compute observables, the one-body density matrix in Eq.~\eqref{rho} is approximated by $\rhoocc$, obtained from Eq.~\eqref{rho} by replacing $U$ with $U_{\rm occ}$, so that
\begin{equation} \label{rhoocc0}
  \rhoocc = (Q \Docc R + I)^{-1} (Q \Docc R)\,.
\end{equation}
This can be reduced to
\begin{equation} \label{rhoocc}
  \rho_{{\rm occ},i j} = \sum_{k,l=1}^{N_{\rm occ}} Q_{i k} (\tilde S + I)^{-1}_{k l} (\Docc R)_{l j} \;,
\end{equation}
which requires $O (N^2_s N_{\mathrm{occ}})$ operations to compute. This scaling follows  from the fact that the inner indices $k,l$ range from $1$ to $N_{\rm occ}$, and by evaluating the expression using matrix multiplications with the appropriate dimensions. Eq.~\eqref{rhoocc} is derived in Appendix A.

\subsection{Canonical-ensemble observables}

To compute the partition function and observables in the canonical ensemble, we diagonalize $\tilde{S}$ (using $O(N_{\rm occ}^3)$ operations),
\begin{equation}\label{tilde-P}
  \tilde{S} = \tilde{P}_S  \tilde\Lambda \tilde{P}_S^{- 1},
\end{equation}
where $\tilde \Lambda_{i j} = \tilde\lambda_{{\rm occ},i} \delta_{i j}$. The eigenvalues  of $S_{\rm occ}$ and hence of $U_{\rm occ} = Q S_{\rm occ} Q^{- 1}$ consist of the $N_{\mathrm{occ}}$ eigenvalues $\tilde\lambda_{{\rm occ},i}$, along with $N_s - N_{\mathrm{occ}}$  zero eigenvalues. Hence, the canonical partition function $Z_N$ can be approximated using Eqs.~{\eqref{Zfourier}} and \eqref{etam}, in which $N_s$ is replaced by $N_{\rm occ}$ and the $N_s$ eigenvalues $\lambda_k$ are replaced by the $N_{\rm occ}$ eigenvalues $\tilde\lambda_{{\rm occ},i}$.

To compute observables in the canonical ensemble, we need the transformation matrices for the eigenvalue decomposition
of $U_{\rm occ} = P_{\rm occ} \Lambda_{\rm occ} P_{\rm occ}^{- 1}$, where $ \Lambda_{\rm occ} $ is a diagonal matrix whose entries are $\Lambda_{{\rm occ},i j} = \lambda_{{\rm occ},i} \, \delta_{i j}\,$, where $\lambda_{{\rm occ},i} = \tilde\lambda_{{\rm occ},i}$ for $i=1,\ldots,\Nocc$ and $\lambda_{{\rm occ},i} = 0$ for $i = \Nocc+1,\ldots,N_s$. 
The matrix elements of $P_{\rm occ}$ and its inverse can be obtained from $\tilde{P}_S$ in \eqref{tilde-P} using $O (N_s N_{\mathrm{occ}}^2)$ operations from the relations
\begin{align}
P_{{\rm occ},i j} & =  \sum_{k = 1}^{N_{\mathrm{occ}}} Q_{i k}  \tilde{P}_{S,
  k j}  \;,\; \;\; i = 1,\ldots, N_s, \quad j = 1,\ldots, N_{\mathrm{occ}} \label{Peqn}\\
 P^{- 1}_{{\rm occ}, i j} & =  \sum_{k = 1}^{N_{\mathrm{occ}}} X^{- 1}_{i k}
 R_{k j} \;,  \;\; \;\;  i = 1,\ldots, N_{\mathrm{occ}}, \quad j = 1,\ldots, N_s , \label{Pinveqn}
\end{align}
where
\begin{equation}\label{X-ik}
  X_{i k} = \sum_{l = 1}^{N_{\mathrm{occ}}} (R Q)_{i l}  \tilde{P}_{S, l k} \;,\qquad i, k = 1,\ldots, N_{\mathrm{occ}} \;. 
\end{equation}
Eqs.~\eqref{Peqn}, \eqref{Pinveqn}, and \eqref{X-ik} are derived in Appendix B. The matrix elements of $P_{\rm occ}$ and $P^{- 1}_{\rm occ}$ that do not appear in Eqs.~{\eqref{Peqn}} and {\eqref{Pinveqn}} do not contribute to observables and may be set to zero.

\begin{figure}[t]
  \begin{center}
  \includegraphics[width=\columnwidth]{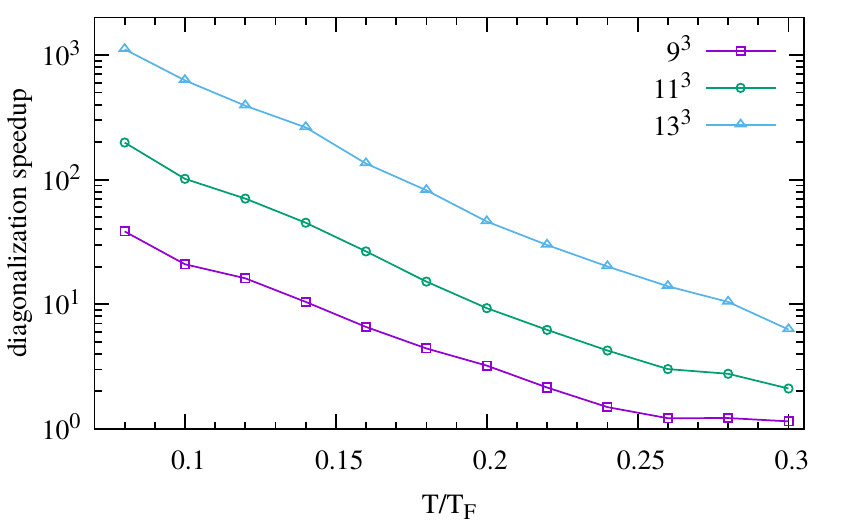}
  \caption{Speedup of the matrix diagonalizations for calculations of the partition function and of observables as a function of $T/T_F$. The speedup shown is computed from the total time spent diagonalizing matrices for a complete sample (averaged over multiple samples), which includes five decorrelation sweeps and the calculation of the energy and heat capacity. The speedup is the ratio of this time using the complete model space of dimension $N_s = N_L^3$ (i.e., by diagonalizing $S$), to the time when using the reduced model space (i.e., by diagonalizing $\tilde S$, the boxed block in Eq.~\eqref{Socc}). Results are shown on a logarithmic scale for the same system as in Fig.~\ref{fig_nred}. \label{fig_diag_speedup}}
  \end{center}
\end{figure}

\begin{figure}[tb]
  \begin{center}
  \includegraphics[width=\columnwidth]{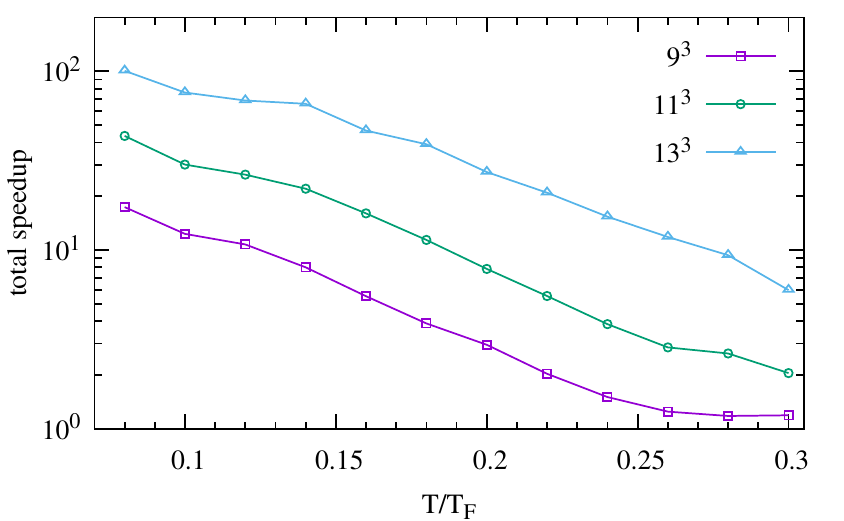}
  \caption{Speedup of the total time per sample as a function of $T/T_F$. The ratio is computed in a similar manner to Fig.~\ref{fig_diag_speedup}, but using the total time per sample including applications of the propagator, QR decompositions, and all other operations needed to evolve the fields and compute observables.  \label{fig_total_speedup}}
  \end{center}
\end{figure}

Canonical expectation values of one-body observables can now be computed using relations similar to Eqs.~(\ref{etam}-\ref{rhoijfourier}) but with only $N_{\mathrm{occ}}$ eigenvalues $\lambda_{{\rm occ},i}$ ($i=1,\ldots,\Nocc$). Because the single-particle model space is reduced to $N_{\mathrm{occ}}$ states, the number of terms in the Fourier sums in Eqs.~(\ref{etam}-\ref{rhoijfourier}) is also reduced to $N_{\mathrm{occ}}$ with quadrature points $\varphi_m = 2 \pi m / N_{\mathrm{occ}}$ for $m = 1, \ldots, N_{\mathrm{occ}}$, and the number of factors in Eq.~\eqref{etam} is reduced to $N_{\mathrm{occ}}$.  In Eq.~\eqref{rhoijfourier} the matrix $P$ is replaced by $P_{\rm occ}$. Computing the one-body density matrix $\rhoocc$ from these equations requires $O(N_s^2 N_{\rm occ})$ operations. 

Note that, as an alternative to the formulas given in Sec.~\ref{gcobs}, grand-canonical expectation values of observables can be also be computed using the canonical-ensemble formulas by taking only one term in the Fourier sums (i.e., setting $N_{\rm FT}=1$ in Eqs.~\eqref{eq:reducedFT} and~\eqref{Znft} below).

Fig.~\ref{fig_diag_speedup} compares the time spent diagonalizing the matrix $S = D R Q$ in a calculation which uses the complete single-particle model space with the time spent diagonalizing $\tilde S$ in a calculation using the reduced space with $\varepsilon = 10^{-6}$. The system is the same as that considered in Fig.~\ref{fig_nred}.
Without the reduction in the single-particle space, the number of operations needed to diagonalize $S$ during an update of all auxiliary fields is $O(N_t N_s^3)$, where $N_s = N_L^3$, and dominates the computational effort of the simulation. With the reduction, the number of operations is $O(N_t N_s N_{\rm occ}^2)$, and the time spent in diagonalization for large lattice sizes is reduced by several orders of magnitude. The number of operations is similarly reduced from $O(N_t N_s^3)$ to  $O(N_t N_s N_{\rm occ}^2)$ in the grand-canonical ensemble.

Fig.~\ref{fig_total_speedup} makes a similar comparison to that in Fig.~\ref{fig_diag_speedup}, but with the total time per sample, including all operations necessary to update the fields and compute observables. As a result of the additional operations, such as QR decomposition and computing time slices, the total speedup is less than that of Fig.~\ref{fig_diag_speedup}, but still can reach multiple orders of magnitude for larger lattices and lower temperatures.

\begin{figure}
  \begin{center}
  \includegraphics[width=\columnwidth]{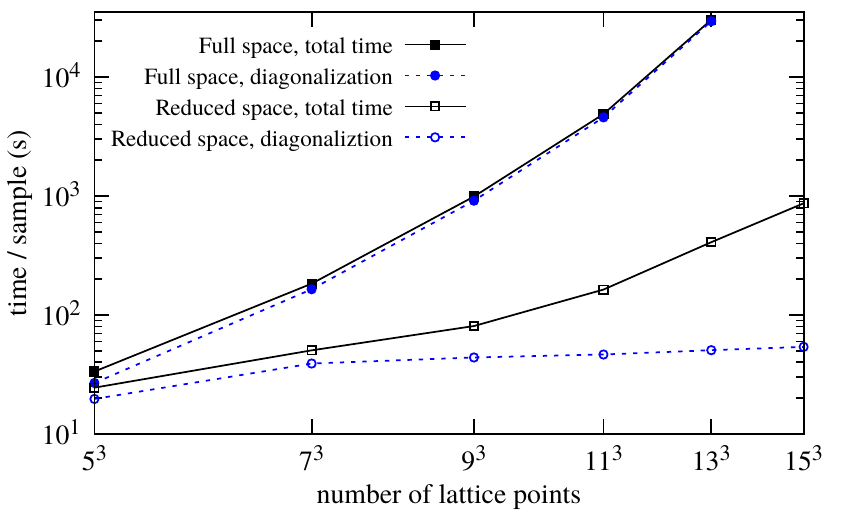}
  \caption{Time (in seconds) per sample on a log scale using the complete single particle model space (solid symbols) and the reduced-dimension single-particle space (open symbols), as a function of lattice size, at $T = 0.10 T_F$. Squares show the total time and circles show the time spent diagonalizing. \label{fig_sample_time}}
  \end{center}
\end{figure}

Fig.~\ref{fig_sample_time} shows the wall-clock time spent diagonalizing, and the total wall-clock time, for the calculation of a single sample with and without the model space reduction for the same calculations as in Fig.~\ref{fig_diag_speedup}. This is shown at single temperature $T/T_F = 0.10$ vs. lattice size. \footnote{The code used for the figures was implemented using MPI + OpenMP parallel processing and Intel MKL routines for matrix operations, with each MPI task corresponding to an independent Monte Carlo walk. For the calculations shown, 8 OpenMP threads were used for each MPI task, and the code was run on the Intel Knight's Landing multicore architecture.}

\subsection{Error estimates}

In this section, we estimate the error that results from omitting the mostly-unoccupied single-particle  states. The grand-canonical partition function $Z = \det (I + U) = \det (I + U_{\rm occ} + E)$ can be expanded in $E$ as
\begin{multline}
    \det (I + U_{\rm occ} + E) = \det (I + U_{\rm occ}) + \\
    \det (I + U_{\rm occ}) \mathrm{tr} [(I + U_{\rm occ})^{- 1} E] + O (E^2)
\end{multline}
Hence the relative error $| Z - Z_{\rm occ} | / | Z |$ in the partition function (where $Z_{\rm occ} = \det (I + U_{\rm occ})$) is given by
\be
(Z - Z_{\rm occ}) / Z \approx \tr [(I - \rho_{\rm occ}) E] \;.
\ee
 The expectation value of a one-body observable $\hat{\mathcal{O}}$ for a given field configuration $\sigma$, computed using $U_{\rm occ}$, acquires an error
\be
\langle \hat{\mathcal{O}} \rangle_\sigma - \langle \hat{\mathcal{O}} \rangle_{\sigma,\mathrm{occ}} = \tr [\mathcal{O} (\rho - \rho_{\rm occ})] \;.
\ee
Expanding $\rho = (I + U)^{-1} U$ where $U = U_{\rm occ} + E$,
\[ (I + U_{\rm occ} + E)^{- 1} = (I + U_{\rm occ})^{-
  1} - (I + U_{\rm occ})^{- 1} E (I + U_{\rm occ})^{- 1} + O (E^2) \;, \]
we find the error in the one-body density $\rho$ to be 
\be
\rho - \rho_{\rm occ} \approx (I - \rho_{\rm occ}) E (I - \rho_{\rm occ}) .
\ee

A general two-body observable $\hat{Y} = \sum_{i j k l} Y_{i j k l} a_i^{\dagger} a_j^{\dagger} a_l a_k $ is characterized by $O(N_s^4)$ non-zero matrix elements, and its expectation value can have a larger error than that of one-body observables.
However, many two-body observables have only $O(N_s)$ or $O(N_s^2)$ nonzero matrix elements. For the observables we have considered  in the example of the unitary Fermi gas, we find that the relative error at a given sample is still generally of order $\varepsilon$.

If $\mu_C$ is determined to approximately give the desired number of particles, the corresponding grand-canonical quantities approximate sufficiently well the canonical expectation values so that the above error estimates also apply for canonical-ensemble observables. As a precaution at low temperatures when there is a well-defined Fermi energy, we can include a certain minimum number of above the Fermi energy regardless of $\varepsilon$.

\begin{table}[ht] 
  \begin{center}
    \begin{tabular}{cccc}
      \hline
      lattice & $Z_N$ (relative)   & $\langle \hat H \rangle_\sigma$ (absolute) & $\langle \hat H \rangle_\sigma$ (scaled) \\ \hline \hline
      $5^3$ & $ 6.3 \times 10^{-8}$ & $ 1.8 \times 10^{-6}$                     & $ 2.6 \times 10^{-8}$  \\
      $7^3$ & $ 1.6 \times 10^{-7}$ & $ 3.4 \times 10^{-6}$ & $ 5.6 \times 10^{-8}$  \\
      $9^3$ & $ 1.3 \times 10^{-7}$ & $ 4.1 \times 10^{-6}$ & $ 6.7 \times 10^{-8}$ \\
      $11^3$ & $ 1.6 \times 10^{-7}$  & $ 4.0 \times 10^{-6}$ & $ 6.7 \times 10^{-8}$ \\
      $13^3$ & $ 1.7 \times 10^{-7}$ & $7.5 \times 10^{-6}$ & $1.3 \times 10^{-7}$\\
      \hline
  \end{tabular}
    \caption{Errors in the partition function $Z_N$ and energy  $\langle \hat H\rangle_\sigma$ due to truncation of the single-particle model space. The errors are computed by performing two separate AFMC calculations, one using the complete model space, and the other using the model space truncation with $\varepsilon = 10^{-6}$, at temperature $T = 0.10 T_F$. All other parameters are the same as in Fig.~\ref{fig_nred}. The $Z_N$ (relative) column lists the maximum relative error $\max_{i} |Z_N^{(i)} - Z_{N,\mathrm{occ}}^{(i)}|/|Z_N^{(i)}|$ for all complete updates of the auxiliary fields $i=1,...,N_{\rm samp} N_{\rm sweep}$ (where $N_{\rm sweep}$ is the number of decorrelation steps per sample). The $\langle \hat H \rangle_\sigma$ (absolute) column lists the maximum absolute error in the energy for all samples, $\max_{n} |\langle \hat
      H \rangle_{\sigma_n} - \langle \hat H \rangle_{\sigma_n,\mathrm{occ}}|$ for $n=1,\ldots,N_{\rm samp}$. The $\langle \hat H \rangle_\sigma$ (scaled) column lists the value of the $\langle \hat H \rangle$ (absolute) column divided by the final expectation $\langle \hat
    H \rangle$ computed from all samples. \label{table_err}}
  \end{center}
\end{table}

Table~\ref{table_err} shows the maximum error in $Z_N$ and $\langle \hat H \rangle_\sigma$ that results from the truncation of the single-particle space for the unitary Fermi gas, using the same parameters as in Fig.~\ref{fig_nred}, at $T = 0.1 T_F$. We find that the sample-by-sample relative error in $Z_N$, and appropriately scaled sample-by-sample absolute error in observables, are of the same order of magnitude as the input parameter $\varepsilon$.

\section{Field updates} \label{sec_field_updates}

To efficiently iterate through and update all time slices $k=1,...,N_t$, we apply a QR decomposition only every $K$ time slices, while still performing a Metropolis update at each time slice. Typically, in order to compute $Z$ or $Z_N$ after updating the fields at imaginary time index $1 < t < N_t$, one needs to compute the product of two $QDR$ decompositions, an $O(N_s^3)$ operation. One can avoid computing such a product at each imaginary time by observing that the factors $U_{N_t} \cdots U_{1}$ can be cyclically permuted while preserving the partition function.

\begin{figure}[h!]
  \begin{center}
  \includegraphics[width=\columnwidth]{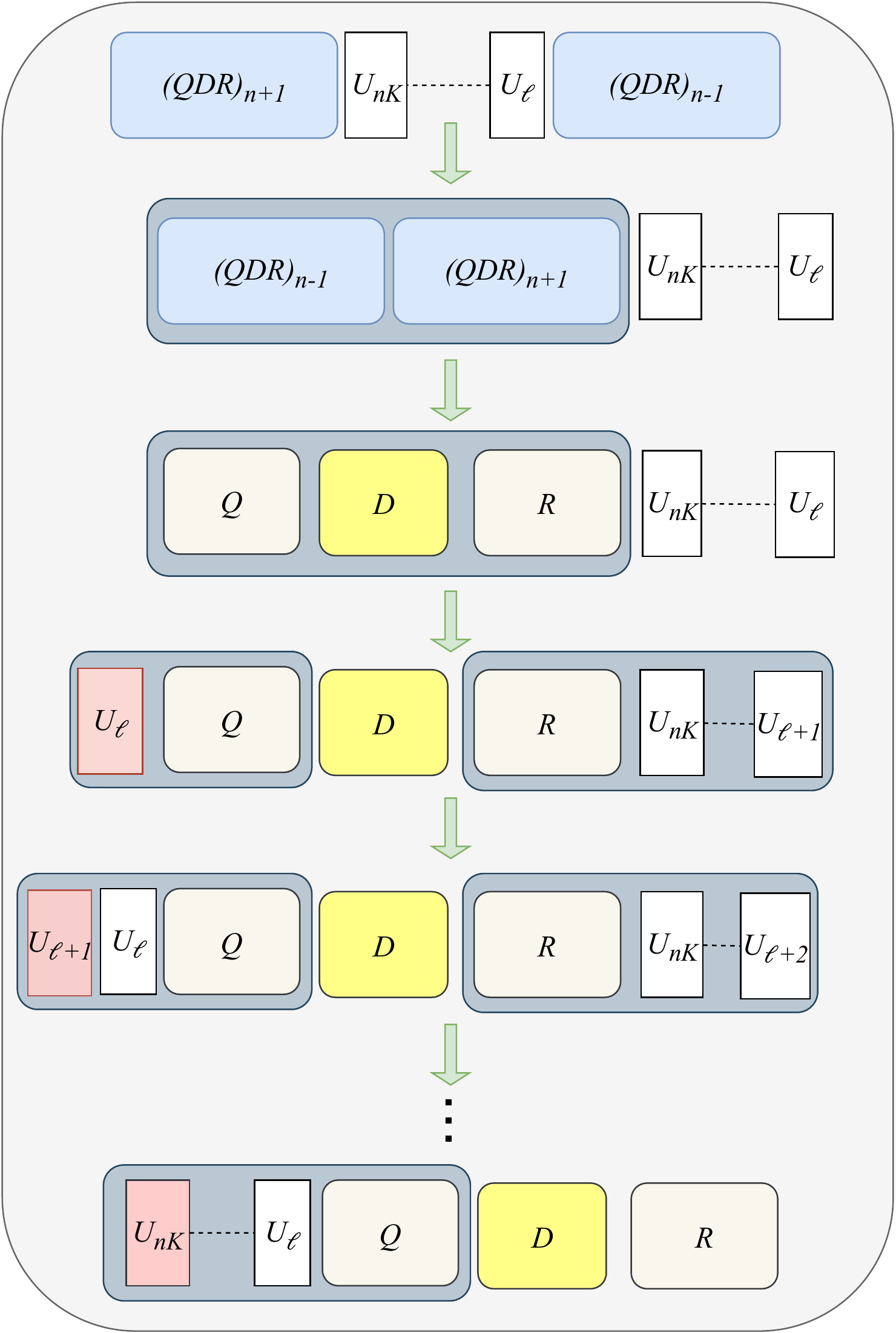}
  \caption{Process of updating the fields while minimizing the number of necesary QR decompositions. See the text for a detailed description. \label{fig_field_updates}}
  \end{center}
\end{figure}

This process is illustrated in Fig.~\ref{fig_field_updates}. Denote by $\ell$ the first element of the $n$th block of $K$ time slices, i.e., $\ell=(n-1) K + 1$. When updating time slice $k$ within this block, $\ell \leq k \leq n K$, the factors in the propagator are grouped as
\begin{align} \label{eq:Ugroup}
  U & = (U_{N_t} \cdots U_{n K+1}) \, (U_{n K} \cdots U_k \cdots U_{\ell}) \, (U_{(n-1)K} \cdots U_1) \nonumber \\
    & = (Q D R)_{n+1} (U_{n K} \cdots U_k \cdots U_{\ell}) (Q D R)_{n-1} \;,
\end{align}
where $(Q D R)_{n+1} = U_{N_t} \cdots U_{n K+1}$ and $(Q D R)_{n-1} = U_{(n-1)K} \cdots U_1$.  The matrix $U$ is related by a similarity transformation to
\begin{equation}
  T \equiv  U_k \cdots U_{\ell} \, (Q D R)_{n-1} (Q D R)_{n+1} U_{n K} \, \cdots U_{k+1} \label{Uprod_sim}\;,
\end{equation}
which conveniently locates $U_k$ as the leftmost factor.

Prior to updating the fields for time slices $k=\ell, \ell+1, \ldots, n K$, we compute a decomposition of the product of $(Q D R)_{n-1}$ and $(Q D R)_{n+1}$, yielding $ Q D R = (Q D R)_{n-1} (Q D R)_{n+1}$, and compute and cache the partial products $R \, U_{n K} \cdots U_k$ for $k={\ell+1}, \ldots, n K$. We then successively update the fields for $k=\ell, {\ell+1}, \ldots, n K$ by accumulating new time slices on the left of the product \eqref{Uprod_sim} and removing old time slices on the right of this product using the precomputed partial products, obtaining for each $k$ a decomposition
\begin{equation}
T = A D B = ( U_k \cdots U_{\ell} Q) D (R \, U_{n K} \cdots U_{k+1}) \, ,
\end{equation}
where $A$ and $B$ are general complex matrices. 

The method of Sec.~\ref{sec_trunc} is used to compute the partition function at time slice $k$ using the decomposition $A D B$, and the new fields at time slice $k$ are accepted or rejected according to the Metropolis algorithm.

It is useful to note that the matrix $T$ is the matrix of the propagator $\hat U(\tau) \hat U(\beta,\tau)$ in the single-particle space. As a result, instead of updating the fields at each $k$ and accepting or rejecting, the decomposition $A D B$ can be used to calculate the force at imaginary time $\tau = k \Delta \beta$ in a determinant-based Hybrid Monte Carlo calculaton.

With this algorithm, QR decompositions are computed only every $K$ time slices, at the cost computing extra applications of time slices. Multiplying by a factor $U_k$ requires $O(N_s^2 \log N_s)$ operations using a fast Fourier transform, if the interaction is local in coordinate space.

Including the major operations, the complexity of the field updates is (for  the Metropolis-Hastings method)
\begin{equation} \label{complexity}
O(N_t N_s^2) + O(N_t N_s N_{\rm occ}^2) + O(N_t N_s^2 \log(N_s)) + O((N_t/K) N_s^3) \;,
\end{equation}
where the first term is from determining $N_{\rm occ}$, the second term is from computing the partition function, the third term is from constructing the time slices, and the fourth term is from the QR decomposition. The benefit of the truncation described in Sec.~\ref{sec_trunc} is to replace the $O(N_t N_s^3)$ complexity of computing the partition function in the complete model space with $O(N_t N_s N_{\rm occ}^2)$. The benefit of this field updating scheme is to reduce the number of required QR decompositions by the factor of $1/K$.

The truncation of the single-particle model space via the $QDR$ decomposition, and the field evolution as described here, was first described in Ref.~\cite{Jensen_review} and was extensively applied in Refs.~\cite{Jensen2020_prl,Jensen_review,Jensen_contact}. At intermediate times $\tau < \beta$, the $QDR$ decomposition could additionally be used to eliminate columns of $Q$ which correspond to very small diagonal elements of $D$ from the field evolution at later times. Thus, in the product $U_{t} \cdots U_{(n-1)K+1} (QDR)_{n-1}$ for $1 < n < N_t/K$, a restricted number of the columns of $Q$ could be considered, which can further reduce the prefactor $N_t/K$ in Eq.~\eqref{complexity}. A method to do this was recently discussed in Ref.~\cite{Shiwei2019}. We postpone further discussion of such an enhancement to a later publication.

\section{Two-body observables}

The expectation value of a two-body observable can be computed from the two-body densities $\langle a_i^{\dagger} a_j a_k^{\dagger} a_l \rangle_\sigma$. In the grand-canonical ensemble, this can be computed from the one-body densities using Wick's theorem
\[ \langle a_i^{\dagger} a_j a_k^{\dagger} a_l \rangle_\sigma = \langle a_i^{\dagger}
   a_j \rangle_\sigma  \langle a_k^{\dagger} a_l \rangle_\sigma + \langle a_i^{\dagger} a_l
   \rangle_\sigma (\delta_{k, j} - \langle a_k^{\dagger} a_j \rangle_\sigma) . \]
In the canonical ensemble
\begin{equation}
  \langle a_i^{\dagger} a_j a_k^{\dagger} a_l \rangle_{\sigma} = \frac{e^{- \beta
  \mu_C}}{Z_N N_s} \sum_{m = 1}^{N_s} e^{- i \varphi_m N} \langle a_i^{\dagger}
  a_j a_k^{\dagger} a_l \rangle_\sigma^{(m)} \eta_m
\end{equation}
where
\begin{equation}
  \langle a_i^{\dagger} a_j a_k^{\dagger} a_l \rangle_\sigma^{(m)} = \gamma_{ji}^{(m)} \gamma_{lk}^{(m)} +
  \gamma_{li}^{(m)} (\delta_{k j} - \gamma_{jk}^{(m)}),
\end{equation}
and $\gamma_{ij}^{(m)} \equiv\langle a_j^\dagger a_i \rangle_\sigma^{(m)}$ are given by Eq.~\eqref{gamma-ij-m}.

In the special case of the two-species cold atom system, in which atoms of the same species do not interact, the two-body canonical-ensemble densities of interest factorize and are simple to compute, i.e., $\langle a_{i \uparrow}^{\dagger} a_{j \uparrow} a_{k \downarrow}^{\dagger} a_{l \downarrow} \rangle_{\sigma} = \langle a_{i \uparrow}^{\dagger} a_{j \uparrow} \rangle_\sigma \langle a_{k \downarrow}^{\dagger} a_{l \downarrow} \rangle_{\sigma}$. This relies on the factorization of the propagator into different species, $\hat{U} = \hat{U}_{\uparrow} \hat{U}_{\downarrow}$ for each auxiliary-field configuration $\sigma$. However, in general this is not the case.


To calculate general two-body observables, it is convenient to first compute the matrices $\gamma_{i j}^{(m)}$. Using the eigenvalue decomposition of $U_{\rm occ}$, we have
\begin{equation}
  \label{gamij} \gamma_{{\rm occ},i j}^{(m)} = \sum_{\alpha = 1}^{N_{\mathrm{occ}}} P_{{\rm occ},i
  \alpha} (1 + \tilde\lambda_{{\rm occ},\alpha}^{- 1} e^{-\beta \mu_C} e^{- i \varphi_m})^{- 1} P^{- 1}_{{\rm occ},\alpha
  j} \;,
\end{equation}
where $\varphi_m = 2 \pi m / N_{\mathrm{occ}}$ for $m = 1, \ldots, N_{\mathrm{occ}}$ and the relevant elements of the matrices $P_{\rm occ}$ are given by Eqs.~\eqref{Peqn}, \eqref{Pinveqn}, and \eqref{X-ik}.  Computing all $N_{\mathrm{occ}}$ matrices $\gamma_{\rm occ}^{(m)}$ in Eq.~{\eqref{gamij}} requires $O (N_s^2 N^2_{\mathrm{occ}})$ operations and $O(N_s^2 N_{\rm occ})$ memory. Without the model space truncation, this calculation requires $O (N_s^4)$ operations and $O (N_s^3)$ memory, both of which become prohibitive for large $N_s$.

\begin{figure}[t]
  \begin{center}
    \includegraphics[width=\columnwidth]{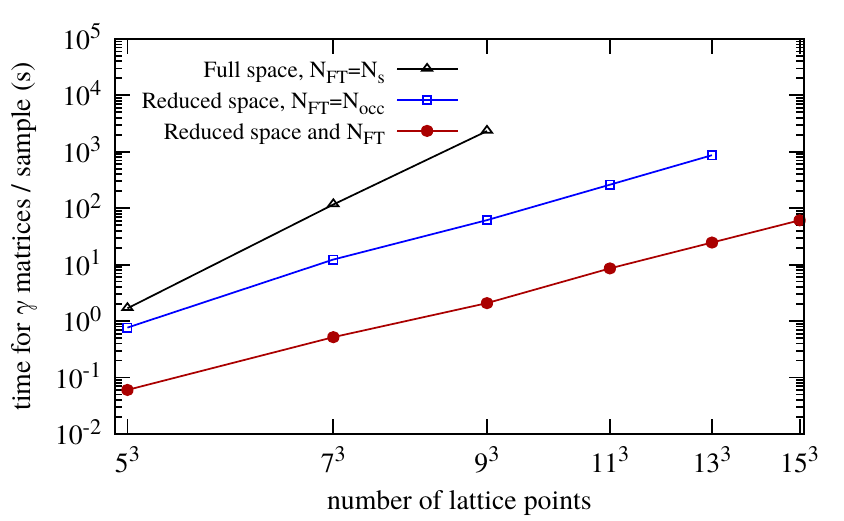}
  \caption{Comparison of the time required to compute all of the matrices $\gamma_{\rm occ}^{(m)}$, on a log scale. The parameters of the calculation are as in Fig.~\ref{fig_sample_time}, except that only one particle-number projection onto the total number of particles $N=66$ is used. Black triangles show the time required using the complete model space of $N_s$ single-particle states and $N_s$ terms in the Fourier transform of the canonical projection. The curve is approximately linear with a slope of $\approx 4$, confirming the $O(N_s^4)$ scaling of the usual two-body canonical projection.  Blue squares show the time required using the reduced model space of $N_{\rm occ}$ states and $N_{\rm FT}=N_{\rm occ}$ terms in the Fourier transform. Purple dots show the time required using the reduced model space of $N_{\rm occ}$ states and a minimal  number of points in the Fourier transform ($N_{\rm FT} \approx 14$). The slope of the later two curves is  $\approx 2$, confirming the reduced $O(N_s^2)$ scaling (note that  $N_{\mathrm{occ}}$ is approximately constant). \label{fig_gamma}}
  \end{center}
\end{figure}

The computational effort and memory required can be reduced further by minimizing the number of quadrature points used in the Fourier sums of  particle-number projected quantities. To accomplish this we extend an idea introduced in Ref.~\cite{Fang2005_thesis} and used in Ref.~\cite{Alhassid2008}. For a given number of quadrature points $N_{\mathrm{FT}}$ in the Fourier sum, we have the identity
\begin{equation} \label{eq:reducedFT}
  \frac{1}{N_{\mathrm{FT}}}  \sum_{m = 1}^{N_{\mathrm{FT}}} e^{i \varphi_m (N -
   N')} = \left\{\begin{array}{ll}
     1\,, & {\rm for} \;\; N = N'  \, (\mathrm{mod}\, N_{\mathrm{FT}})\\
     0\,, & \mathrm{otherwise}
  \end{array}\right.,
  \end{equation}
where $\varphi_m = 2 \pi m / N_{\mathrm{FT}}$. Using $N_{\mathrm{FT}}$ quadrature points to calculate the canonical partition function, we obtain
\begin{eqnarray}
  Z_N (N_{\mathrm{FT}}) & \equiv & \frac{e^{- \beta \mu_C
  N}}{N_{\mathrm{FT}}}  \sum_{m = 1}^{N_{\mathrm{FT}}} e^{- i \varphi_m N}
  \Tr (e^{i \varphi_m  \hat{N}} e^{\beta \mu_C \hat{N}} \hat{U}) \nonumber \\
  & = & \sum_{k=0,\pm 1,\ldots} \Tr_{N + k N_{\mathrm{FT}}} (\hat{U} e^{\beta \mu_C (\hat{N} - N)}) \label{Znft} .
\end{eqnarray}
The reduced number of quadrature points therefore includes contributions from particle numbers $N$, $N \pm N_{\mathrm{FT}}$, $N \pm 2 N_{\mathrm{FT}}$, $\ldots$.  With $\mu_C$ chosen to give  $\langle \hat N\rangle_\sigma \approx N$, the largest term on the r.h.s.~of Eq.~{\eqref{Znft}} is the $N$-particle trace, with the $N \pm N_{\mathrm{FT}}$, $N \pm 2 N_{\mathrm{FT}}$, ... traces giving increasingly smaller corrections.

Given a desired error tolerance $\varepsilon$ for the particle-number projection, we can determine the minimal necessary value of  $N_{\mathrm{FT}}$ by starting with a small value of $N_{\mathrm{FT}}$ and computing the partition function $Z_N$ and diagonal one-body densities $\rho_{\alpha}$ for increasingly large values of $N_{\mathrm{FT}}$ until the conditions
\begin{align} 
  | \log Z_N - \log Z_N (N_{\mathrm{FT}}) | & < \varepsilon\;, \label{NFT_cond_Z} \\
  \sum_\alpha | \rho_{\alpha} - \rho_{\alpha}(N_{\mathrm{FT}}) | & < \varepsilon \label{NFT_cond_rho}\;,
\end{align}
are both satisfied. Although this $N_{\mathrm{FT}}$ is determined by the one-body densities, it can in practice be effectively used also for two-body observables. Using this value of $N_{\rm FT}$, the calculation of the $\gamma_{\rm occ}^{(m)}$ matrices in Eq.~{\eqref{gamij}} requires $O (N_s^2 N_{\mathrm{occ}} N_{\mathrm{FT}})$ operations and $O(N_s^2 N_{\rm FT})$ memory.

The contributions from fully occupied states with $(1 + \lambda_{{\rm occ},\alpha}^{- 1} e^{- i \varphi_m})^{- 1} \approx 1$ in Eq.  {\eqref{gamij}} are independent of $m$, and thus can be calculated separately, offering an additional optimization.

Fig.~\ref{fig_gamma} shows a comparison of the time required to compute all of the matrices $\gamma_{\rm occ}^{(m)}$ with and without the truncation in the size of the single-particle model space and the reduction in the number of points $N_{\mathrm{FT}}$ in the Fourier transform of the particle-number projection. Here we used a single particle-number projection onto the total number of particles $N=66$, as done for our calculations for the spin susceptibility in Ref.~\cite{Jensen2020_prl}. We find $N_{FT} \approx 14$, independent of the lattice size. The results in Fig.~\ref{fig_gamma} (shown on a logarithmic scale) confirms the reduction of the $O(N_s^3)$ scaling  (slope of $\approx 3$) when using the full single-particle model space to $O(N_s^2)$ when using the reduced model space (slope of $\approx 2$).

\section{Summary}

Below we summarize the methods described in this work.

\emph{Single-particle model space reduction.} \, Given an auxiliary field configuration $\sigma$ and a decomposition $Q D R$ of the matrix $U$ representing the propagator (or a matrix related to $U$ by a similarity transformation), the number $\Nocc$ of significantly occupied states is determined using Eq.~\eqref{enrm}, given an input parameter $\varepsilon$ specifying the approximate error tolerance for observables. The diagonal elements of $D$ corresponding to mostly unoccupied states are set to zero, and hence are omitted from all subsequent matrix operations.

\emph{Partition function and observables.} \, The partition function and observables are computed using the space of reduced dimension $\Nocc$ by constructing the matrix $\tilde S$, which is the boxed block in Eq.~\eqref{Socc}. In the grand-canonical ensemble, the partition function and observables are computed using Eqs.~\eqref{Zocc} and~\eqref{rhoocc}. In the canonical ensemble, $\tilde S$ is diagonalized to obtain an eigenvalue decomposition of $\Uocc$ using Eqs.~(\ref{tilde-P}-\ref{X-ik}). The $\Nocc$ eigenvalues $\tilde \lambda_{{\rm occ},i}$ of $\tilde S$ are the only nonzero eigenvalues of $\Uocc$. The partition function and observables are computed, respectively, using Eq.~(\ref{Zfourier}) and  Eqs.~\eqref{etam}-\eqref{rhoijfourier}, replacing $N_{s}$ with $N_{\rm occ}$, $\varphi_m$ with $\varphi_m = 2\pi m/\Nocc$ ($m=1,\ldots,\Nocc$), $\lambda_i$ with $\tilde \lambda_{{\rm occ},i}$, and the matrix $P$ with $P_{\rm occ}$.

The error in these observables due to the truncation is controlled by $\varepsilon$ and can be made arbitrarily small, at the expense of $\Nocc$ becoming larger.


\emph{Field updates.} \, To avoid calculation of QR decompositions at each imaginary time while updating the fields, the time slices may be cyclically permuted according to the method of Sec.~\ref{sec_field_updates} to produce a matrix which differs from $U$ by a similarity transformation. The partition function for each trial configuration is calculated using the reduced model space according to the method just described.

\emph{Canonical two-body observables.} \, Canonical two-body observables are calculated using the reduced model space and an additionally reduced number of quadrature points $N_{\rm FT} \leq \Nocc$ in the Fourier sum. The number of points $N_{\rm FT}$ is determined using Eqs.~(\ref{NFT_cond_Z},\ref{NFT_cond_rho}), again with a controlled error determined by an input parameter $\varepsilon$. The $\gamma^{(m)}$ matrices ($m=1,\ldots,N_{\rm FT}$)  are calculated from Eq.~\eqref{gamij}, with $\varphi_m = 2 \pi m/N_{\rm FT}$.

\section{Conclusion}

Finite-temperature auxiliary-field quantum Monte Carlo (AFMC) calculations typically have a computational scaling of $O (N_s^3)$ or higher power in $N_s$, where $N_s$ is the number of single-particle states. We have introduced an improvement which significantly reduces the computational effort by eliminating states which are mostly unoccupied from the calculation of the partition function and observables at any given configuration of the auxiliary fields. Since the single-particle states must be periodically re-orthogonalized as the single-particle propagator $\hat U(\sigma)$ is applied, this produces a decomposition (such as the $Q D R$ decomposition), which allows us to identify and eliminate these states.  The remaining model space has a reduced dimension $N_{\mathrm{occ}}$ of significantly occupied states, which scales gently with the number of particles $N$ and is usually of the order $N$. Our method is applicable in both canonical ensemble and grand-canonical ensemble calculations.  The speedup of our algorithm is particularly substantial in systems with $N \ll N_s $. 

For the canonical ensemble, the calculation of general two-body observables in the reduced model space would require the calculation of $N_{\rm occ}$ matrices of dimension $N_s$ if the usual particle-number projection were used. We have improved the particle-number projection method to further reduce the required number of these matrices, greatly reducing the required memory and number of operations for canonical-ensemble two-body observables.

These improvements can enable otherwise impractical studies of dilute fermionic systems.  In particular, we have used these methods to carry out finite-temperature lattice AFMC calculations for the spin-balanced unitary Fermi gas (where $N_s$ is proportional to the cubic power of the linear size of the lattice), including precision studies of the pseudogap phenomenon~\cite{Jensen2020_prl,Jensen_review}, and of thermodynamic observables in the continuum limit~\cite{Jensen_contact}.

\section*{Acknowledgments}

This work was supported in part by the U.S.~DOE grants Nos.~DE-SC0019521, DE-FG02-91ER40608, and DE-FG02-00ER41132. 
The research presented here used resources of the National Energy Research Scientific Computing Center, which is supported by the Office of Science of the U.S.~Department of Energy under Contract No.~DE-AC02-05CH11231.  We also thank the Yale Center for Research Computing for guidance and use of the research computing infrastructure.

\section*{Appendix A}
\label{secAppendixA}

In this Appendix we derive Eq.~\eqref{rhoocc} from the expresssion $\rhoocc = (Q \Docc R + I)^{-1} (Q \Docc R)$.

First, we have
\begin{align}
  (Q \Docc R + I)^{-1} & = Q (\Docc R Q  + I)^{-1} Q^{-1} \,,
\end{align}
and hence
\begin{align}
  \rhoocc & = Q (\Docc R Q  + I)^{-1} \Docc R \;.
\end{align}
The matrix $\Docc R Q + I$ has the form
\[
\Docc R Q + I =
\left(
\begin{array}{@{\extracolsep{3pt}}ccccll}
\cline{1-3} \cline{4-6}
\multicolumn{3}{|c|}{}  & \multicolumn{3}{|c|}{\hspace{1cm}} \\
\multicolumn{3}{|c|}{\tilde{S} + I_{\Nocc}} & \multicolumn{3}{|c|}{Y} \\
\multicolumn{3}{|c|}{} & \multicolumn{3}{|c|}{} \\
\cline{1-3} \cline{4-6}
0 & 0 & 0 & \multicolumn{3}{|c|}{\multirow{2}{1cm}{$I_{N_s-\Nocc}$}} \\
0 & 0 & 0 & \multicolumn{3}{|c|}{} \\ \cline{4-6} 
\end{array}
\right), \]
where $I_n$ is the $n \times n$ identity matrix and $Y$ is the upper right $\Nocc \times (N_s-\Nocc)$ block. Hence its inverse has the form
\[
(\Docc R Q + I)^{-1} =
\left(
\begin{array}{@{\extracolsep{3pt}}ccccll}
\cline{1-3} \cline{4-6}
\multicolumn{3}{|c|}{}  & \multicolumn{3}{|c|}{\hspace{1cm}} \\
\multicolumn{3}{|c|}{(\tilde{S} + I_{\Nocc})^{-1}} & \multicolumn{3}{|c|}{Z} \\
\multicolumn{3}{|c|}{} & \multicolumn{3}{|c|}{} \\
\cline{1-3} \cline{4-6}
0 & \;\; 0 & 0 & \multicolumn{3}{|c|}{\multirow{2}{1cm}{$I_{N_s-\Nocc}$}} \\
0 & \;\; 0 & 0 & \multicolumn{3}{|c|}{} \\ \cline{4-6} 
\end{array}
\right), \]
where $Z$ is an $\Nocc \times (N_s-\Nocc)$ matrix satisfying $(\tilde S + I_{\Nocc})Z + Y = 0$, although $Z$ is not needed for our calculation.

Since the only nonzero entries of $\Docc$ are its first $\Nocc$ diagonal elements, $(\Docc R Q + I)^{-1} \Docc$ has the form
\[
(\Docc R Q + I)^{-1} \Docc =
\left(
\begin{array}{@{\extracolsep{3pt}}ccccl}
\cline{1-3}
\multicolumn{3}{|c|}{} & 0 & 0 \\
\multicolumn{3}{|c|}{(\tilde{S} + I_{\Nocc})^{-1} \tilde{D}_{\rm occ}} & 0 & 0  \\
\multicolumn{3}{|c|}{} & 0 & 0  \\
\cline{1-3}
0 & \;\;\;\; 0 & \;\;\;\; 0 & 0 & 0  \\
0 & \;\;\;\; 0 & \;\;\;\; 0 & 0 & 0  \\
\end{array}
\right), \]
where $\tilde{D}_{\rm occ}$ is the upper left $\Nocc \times \Nocc$ block of $\Docc$; i.e., all of the entries of $(\Docc R Q + I)^{-1} \Docc$ are zero except its upper left $\Nocc \times \Nocc$ block, which can be determined by inverting the $\Nocc \times \Nocc$ matrix $\tilde{S} + I_{\Nocc}$. Eq.~\eqref{rhoocc} therefore follows, allowing $\rhoocc$ to be calculated using $O(N_s^2 \Nocc)$ operations.
\section*{Appendix B}

\label{secAppendixB}

In this Appendix we discuss how to obtain the eigenvectors and eigenvalues of $S_{\rm occ}$ from those of the truncated matrix $\tilde{S}$.  Since $\Uocc$ is obtained from $S_{\rm occ}$ by a similarity transformation, this will provide us with the eigenvector decomposition of $\Uocc$. 

Consider a nonzero eigenvalue $\lambda$ of $S_{\rm occ}$. If $S_{\rm occ} \, \bm{y} = \lambda \bm{y}$, then since the last $N_s - N_{\rm occ}$ rows of $S_{\rm occ}$ are zero, so are the corresponding entries of $\lambda \bm{y}$. If $\lambda \neq 0$, this means the last $N_s - N_{\mathrm{occ}}$ entries of $\bm{y}$ are zero. Hence all right eigenvectors of $S_{\rm occ}$ with $\lambda \neq 0$ have the form
\[ \bm{y} = \left(\begin{array}{c}
     y_1\\
     \vdots\\
     y_{N_{\mathrm{occ}}}\\
     0\\
     \vdots
   \end{array}\right) \;\; {\rm for} \;\; \lambda \neq 0\;, \]
i.e., are contained in the subspace $V_{N_{\rm occ}} = \mathrm{sp} \{\bm{e}_1, \ldots, \bm{e}_{N_{\mathrm{occ}}} \}$ spanned by the first $N_{\mathrm{occ}}$ unit vectors $\bm{e}_i$. Since $S_{\rm occ}$ maps this space onto itself, $S_{\rm occ} : V_{N_{\rm occ}} \rightarrow V_{N_{\rm occ}}$, one can obtain all right eigenvectors with non-zero eigenvalues by diagonalizing $S_{\rm occ}$ in this subspace. The matrix of $S_{\rm occ}$ in this subspace is the $N_{\rm occ}\times N_{\rm occ}$ matrix $\tilde{S}$.

Any right eigenvector of $\tilde{S}$ therefore corresponds to a right eigenvector of $S_{\rm occ}$, obtained by appending zeros. The other right eigenvectors of $S_{\rm occ}$ have $\lambda = 0$ and are exactly the elements of the nullspace of $S_{\rm occ}$. Since the last $N_s - N_{\mathrm{occ}}$ rows of $S_{\rm occ}$ are zero, the nullspace is
\begin{multline}
  \label{null} \mathrm{Null} (S_{\rm occ}) = \{\bm{x} \in V \;|\; \bm{x} \text{ is orthogonal to} \\
  \text{the first $N_{\mathrm{occ}}$ rows of $S_{\rm occ}$} \} .
\end{multline}
Note generally $\mathrm{Null} (S_{\rm occ}) \neq \mathrm{sp}
\{\bm{e}_{N_{\mathrm{occ}} + 1}, \ldots, \bm{e}_{N_s} \}$. So the transformation matrix $P_S$ in the decomposition $S_{\rm occ} = P_S \Lambda_{\rm occ} P_S^{-1}$ has the form
\[ P_S =
\left(
\begin{array}{@{\extracolsep{-4pt}}ccccc}
  \Big\vert &  & \Big\vert  & \Big\vert & \\[6pt]
     \bm{y}_1 & \ldots & \bm{y}_{N_{\rm occ}} & \bm{n}_1 & \ldots \\[6pt]
     \Big\vert &  & \Big\vert & \Big\vert &\\
\end{array}
\right) =
\left(
\begin{array}{@{\extracolsep{3pt}}ccccll}
\cline{1-3} \cline{4-6}
\multicolumn{3}{|c|}{}  & \multicolumn{3}{|c|}{\multirow{5}{1.5cm}{Basis for $\mathrm{Null}(S_{\mathrm{occ}})$}} \\
\multicolumn{3}{|c|}{ \tilde{P}_S } & \multicolumn{3}{|c|}{} \\
\multicolumn{3}{|c|}{} & \multicolumn{3}{|c|}{} \\ \cline{1-3}
0 & 0 & 0 & \multicolumn{3}{|c|}{} \\
0 & 0 & 0 & \multicolumn{3}{|c|}{} \\ \cline{4-6} 
\end{array}
\right), \] where $\tilde{P}_S$ is the matrix whose columns are the eigenvectors of $\tilde{S}$. We need only the first $N_{\mathrm{occ}}$ columns of $P_S$ for computing the density matrix.

We also need the first $N_{\mathrm{occ}}$ rows of the inverse $P_S^{-1}$, i.e., the matrix whose rows are the left eigenvectors $\bm{w}_i$ of $S_{\rm occ}$. This has the form
\[ P_S^{- 1} = \left(\begin{array}{c}
     \rule{.7cm}{.4pt} \;\;\;\; \bm{w}_1 \;\;\;\;\rule{.7cm}{.4pt}\\[-2pt]
     \vdots\\[-2pt]
     \rule{.7cm}{.4pt} \;\; \bm{w}_{N_{\mathrm{occ}}} \;\;\rule{.7cm}{.4pt}\\[2pt]
     \vdots\\[3pt]
\end{array}
\right) =
\left(
\begin{array}{@{\extracolsep{3pt}}ccccc}
\cline{1-3} \cline{4-5}
\multicolumn{3}{|c|}{}  & \multicolumn{2}{|c|}{} \\[2pt]
\multicolumn{3}{|c|}{ \tilde{P}_S^{-1} } & \multicolumn{2}{|c|}{A} \\[2pt]
\multicolumn{3}{|c|}{} & \multicolumn{2}{|c|}{} \\[2pt]\cline{1-3} \cline{4-5}
x & x & x & x & x \\
x & x & x & x & x \\
\end{array}
\right) \;.
\]
The row vectors $\bm{w}_i$ for $i = 1, \ldots, N_{\mathrm{occ}}$ are determined by the condition $P^{-1}_S P_S = I$, i.e.,
\begin{equation}
  \label{dotcond} \bm{w}_i \cdot \bm{y}_j = \delta_{i, j} \,, \quad \bm{w}_i \cdot \bm{n}_k = 0
  \qquad (i, j = 1, \ldots, N_{\rm occ} ; \: \mathrm{all} \, k) \,,
\end{equation}
where $\bm{n}_k, \; k = 1, \ldots, N_s - N_{\mathrm{occ}}$ is any basis for $\mathrm{Null} (S_{\rm occ})$. The first $N_{\mathrm{occ}}$ entries of the vector $\bm{w}_i$ are the entries in the $i$th row of $\tilde{P}_S^{- 1}$, since one has the identity $\tilde{P}_S^{- 1} \tilde{P}_S = I$ for the block sub-matrices $\tilde{P}_S, \tilde{P}_S^{- 1}$. However, the remaining $N_s - N_{\mathrm{occ}}$ entries are generally nonzero so that $\bm{w}_i$ can be orthogonal to all of the $\bm{n}_k$. These entries comprise the submatrix $A$, which is unknown and must be determined.

To determine $A$, consider that since $\bm{w}_i$ must be orthogonal to the nullspace of $S_{\rm occ}$, by Eq. {\eqref{null}} it is contained in the span of the first $N_{\mathrm{occ}}$ rows $\bm{r}_k$ of $R Q$. We can write
\begin{equation}
  \label{wcomb} \bm{w}_i = \sum_{k = 1}^{N_{\mathrm{occ}}} B_{i k} \bm{r}_k  \,,
\end{equation}
for some coefficients $B_{i k}$. Then the first part of condition
{\eqref{dotcond}} becomes
\[ \bm{w}_i \cdot \bm{y}_j = \sum_{k = 1}^{N_{\mathrm{occ}}} B_{i k} \bm{r}_k \cdot \bm{y}_j =
   \delta_{i j} \,, \]
so we obtain
\[ B_{i k} = X^{- 1}_{i k} \quad \text{where} \quad X_{i k} \equiv \bm{r}_i \cdot
\bm{y}_k \: . \]
The first $N_{\mathrm{occ}}$ rows of $P_S^{- 1}$ can therefore be obtained by inverting the $N_{\mathrm{occ}} \times N_{\mathrm{occ}}$ matrix $X$ and forming the linear combinations {\eqref{wcomb}}.
In terms of matrix operations,
\begin{equation}
P^{- 1}_{S, i j} = \sum_{k = 1}^{N_{\mathrm{occ}}} X^{- 1}_{i k}  (R Q)_{k
  j}, \quad  X_{i k} = \sum_{l = 1}^{N_{\mathrm{occ}}} (R Q)_{i l} P_{S, l k} \:,
\end{equation}
where $i, k = 1, \ldots, N_{\mathrm{occ}}$ and $j = 1, \ldots, N_s$ .

Now, to obtain the eigenvalue decomposition $U_{\rm occ} = P_{\rm occ} \Lambda_{\rm occ} P_{\rm occ}^{- 1}$, we observe that $U_{\rm occ} = Q S_{\rm occ} Q^{- 1} = (Q P_S ) \Lambda_{\rm occ} (P_S^{- 1} Q^{- 1})$, and obtain
\begin{equation}
  \label{Peqnappdx} P_{{\rm occ},i j} = \sum_{k = 1}^{N_{\mathrm{occ}}} Q_{i k} 
  \tilde{P}_{S, k j}, \qquad i = 1, \ldots, N_s, \quad j = 1, \ldots, N_{\mathrm{occ}}
\end{equation}
\begin{equation}
  \label{Pinveqnappdx} P^{- 1}_{{\rm occ},i j} = \sum_{k = 1}^{N_{\mathrm{occ}}} X^{-
  1}_{i k} R_{k j}, \qquad i = 1, \ldots, N_{\mathrm{occ}}, \quad j = 1, \ldots, N_s 
\end{equation}
where Eqs. {\eqref{Peqnappdx}} and {\eqref{Pinveqnappdx}} specify only the indices $i, j$ which contribute to observables. The other matrix elements may be set to zero.

\bibliographystyle{apsrev}
\bibliography{reducing}

\begin{thebibliography}{27}
\expandafter\ifx\csname natexlab\endcsname\relax\def\natexlab#1{#1}\fi
\expandafter\ifx\csname bibnamefont\endcsname\relax
  \def\bibnamefont#1{#1}\fi
\expandafter\ifx\csname bibfnamefont\endcsname\relax
  \def\bibfnamefont#1{#1}\fi
\expandafter\ifx\csname citenamefont\endcsname\relax
  \def\citenamefont#1{#1}\fi
\expandafter\ifx\csname url\endcsname\relax
  \def\url#1{\texttt{#1}}\fi
\expandafter\ifx\csname urlprefix\endcsname\relax\def\urlprefix{URL }\fi
\providecommand{\bibinfo}[2]{#2}
\providecommand{\eprint}[2][]{\url{#2}}

\bibitem[{\citenamefont{Jensen et~al.}(2019)\citenamefont{Jensen, Gilbreth, and
  Alhassid}}]{Jensen_review}
\bibinfo{author}{\bibfnamefont{S.}~\bibnamefont{Jensen}},
  \bibinfo{author}{\bibfnamefont{C.~N.} \bibnamefont{Gilbreth}},
  \bibnamefont{and} \bibinfo{author}{\bibfnamefont{Y.}~\bibnamefont{Alhassid}},
  \bibinfo{journal}{Eur. Phys. J. Spec. Top.} \textbf{\bibinfo{volume}{227}},
  \bibinfo{pages}{2241} (\bibinfo{year}{2019}).

\bibitem[{\citenamefont{Jensen et~al.}(2020{\natexlab{a}})\citenamefont{Jensen,
  Gilbreth, and Alhassid}}]{Jensen2020_prl}
\bibinfo{author}{\bibfnamefont{S.}~\bibnamefont{Jensen}},
  \bibinfo{author}{\bibfnamefont{C.~N.} \bibnamefont{Gilbreth}},
  \bibnamefont{and} \bibinfo{author}{\bibfnamefont{Y.}~\bibnamefont{Alhassid}},
  \bibinfo{journal}{Phys. Rev. Lett.} \textbf{\bibinfo{volume}{124}},
  \bibinfo{pages}{090604} (\bibinfo{year}{2020}{\natexlab{a}}).

\bibitem[{\citenamefont{Jensen et~al.}(2020{\natexlab{b}})\citenamefont{Jensen,
  Gilbreth, and Alhassid}}]{Jensen_contact}
\bibinfo{author}{\bibfnamefont{S.}~\bibnamefont{Jensen}},
  \bibinfo{author}{\bibfnamefont{C.~N.} \bibnamefont{Gilbreth}},
  \bibnamefont{and} \bibinfo{author}{\bibfnamefont{Y.}~\bibnamefont{Alhassid}},
  \bibinfo{journal}{Phys. Rev. Lett.} \textbf{\bibinfo{volume}{125}},
  \bibinfo{pages}{043402} (\bibinfo{year}{2020}{\natexlab{b}}).

\bibitem[{\citenamefont{Koonin et~al.}(1997)\citenamefont{Koonin, Dean, and
  Langanke}}]{Koonin1997_physrep}
\bibinfo{author}{\bibfnamefont{S.}~\bibnamefont{Koonin}},
  \bibinfo{author}{\bibfnamefont{D.}~\bibnamefont{Dean}}, \bibnamefont{and}
  \bibinfo{author}{\bibfnamefont{K.}~\bibnamefont{Langanke}},
  \bibinfo{journal}{Phys. Rep.} \textbf{\bibinfo{volume}{278}},
  \bibinfo{pages}{2 } (\bibinfo{year}{1997}), ISSN \bibinfo{issn}{0370-1573}.

\bibitem[{\citenamefont{Alhassid}(2001)}]{Alhassid2001_ijmpb}
\bibinfo{author}{\bibfnamefont{Y.}~\bibnamefont{Alhassid}},
  \bibinfo{journal}{Int. J. Mod. Phys. B} \textbf{\bibinfo{volume}{15}},
  \bibinfo{pages}{1447} (\bibinfo{year}{2001}).

\bibitem[{\citenamefont{Alhassid}(2017)}]{Alhassid2017}
\bibinfo{author}{\bibfnamefont{Y.}~\bibnamefont{Alhassid}}, in
  \emph{\bibinfo{booktitle}{Emergent Phenomena in Atomic Nuclei from
  Large-Scale Modeling: a Symmetry-Guided Perspective}}, edited by
  \bibinfo{editor}{\bibfnamefont{K.~D.} \bibnamefont{Launey}}
  (\bibinfo{publisher}{World Scientific}, \bibinfo{address}{Singapore},
  \bibinfo{year}{2017}), pp. \bibinfo{pages}{267--298}.

\bibitem[{\citenamefont{Hubbard}(1959)}]{Hubbard1959}
\bibinfo{author}{\bibfnamefont{J.}~\bibnamefont{Hubbard}},
  \bibinfo{journal}{Phys. Rev. Lett.} \textbf{\bibinfo{volume}{3}},
  \bibinfo{pages}{77} (\bibinfo{year}{1959}).

\bibitem[{\citenamefont{Stratonovich}(1957)}]{Stratonovich1957}
\bibinfo{author}{\bibfnamefont{R.}~\bibnamefont{Stratonovich}},
  \bibinfo{journal}{Dokl. Akad. Nauk. S.S.S.R.} \textbf{\bibinfo{volume}{115}},
  \bibinfo{pages}{1097} (\bibinfo{year}{1957}), \bibinfo{note}{[Sov. Phys.
  Dokl. \textbf{2}, 416 (1957)]}.

\bibitem[{\citenamefont{Alhassid et~al.}(2007)\citenamefont{Alhassid, Liu, and
  Nakada}}]{Alhassid2007}
\bibinfo{author}{\bibfnamefont{Y.}~\bibnamefont{Alhassid}},
  \bibinfo{author}{\bibfnamefont{S.}~\bibnamefont{Liu}}, \bibnamefont{and}
  \bibinfo{author}{\bibfnamefont{H.}~\bibnamefont{Nakada}},
  \bibinfo{journal}{Phys. Rev. Lett.} \textbf{\bibinfo{volume}{99}},
  \bibinfo{pages}{162504} (\bibinfo{year}{2007}).

\bibitem[{\citenamefont{Beyl et~al.}(2018)\citenamefont{Beyl, Goth, and
  Assaad}}]{Assaad2018_HMC}
\bibinfo{author}{\bibfnamefont{S.}~\bibnamefont{Beyl}},
  \bibinfo{author}{\bibfnamefont{F.}~\bibnamefont{Goth}}, \bibnamefont{and}
  \bibinfo{author}{\bibfnamefont{F.~F.} \bibnamefont{Assaad}},
  \bibinfo{journal}{Phys. Rev. B} \textbf{\bibinfo{volume}{97}},
  \bibinfo{pages}{085144} (\bibinfo{year}{2018}).

\bibitem[{\citenamefont{Duane et~al.}(1987)\citenamefont{Duane, Kennedy,
  Pendleton, and Roweth}}]{Duane1987_HMC}
\bibinfo{author}{\bibfnamefont{S.}~\bibnamefont{Duane}},
  \bibinfo{author}{\bibfnamefont{A.}~\bibnamefont{Kennedy}},
  \bibinfo{author}{\bibfnamefont{B.~J.} \bibnamefont{Pendleton}},
  \bibnamefont{and} \bibinfo{author}{\bibfnamefont{D.}~\bibnamefont{Roweth}},
  \bibinfo{journal}{Phys. Lett. B} \textbf{\bibinfo{volume}{195}},
  \bibinfo{pages}{216 } (\bibinfo{year}{1987}), ISSN \bibinfo{issn}{0370-2693}.

\bibitem[{\citenamefont{Kennedy}(2007)}]{Kennedy_qcd}
\bibinfo{author}{\bibfnamefont{A.~D.} \bibnamefont{Kennedy}}, in
  \emph{\bibinfo{booktitle}{Perspectives in Lattice QCD}}
  (\bibinfo{publisher}{World Scientific}, \bibinfo{year}{2007}).

\bibitem[{\citenamefont{Alhassid}(2013)}]{Alhassid2013}
\bibinfo{author}{\bibfnamefont{Y.}~\bibnamefont{Alhassid}}, in
  \emph{\bibinfo{booktitle}{Fifty Years of Nuclear BCS: Pairing in Finite
  Systems}}, edited by
  \bibinfo{editor}{\bibfnamefont{R.}~\bibnamefont{Broglia}} \bibnamefont{and}
  \bibinfo{editor}{\bibfnamefont{V.}~\bibnamefont{Zelevinsky}}
  (\bibinfo{publisher}{World Scientific}, \bibinfo{address}{Singapore},
  \bibinfo{year}{2013}), pp. \bibinfo{pages}{608--626}.

\bibitem[{\citenamefont{Gilbreth and Alhassid}(2013)}]{Gilbreth2013_pra}
\bibinfo{author}{\bibfnamefont{C.~N.} \bibnamefont{Gilbreth}} \bibnamefont{and}
  \bibinfo{author}{\bibfnamefont{Y.}~\bibnamefont{Alhassid}},
  \bibinfo{journal}{Phys. Rev. A} \textbf{\bibinfo{volume}{88}},
  \bibinfo{pages}{063643} (\bibinfo{year}{2013}).

\bibitem[{\citenamefont{Ormand et~al.}(1994)\citenamefont{Ormand, Dean,
  Johnson, Lang, and Koonin}}]{Ormand1994_projection}
\bibinfo{author}{\bibfnamefont{W.~E.} \bibnamefont{Ormand}},
  \bibinfo{author}{\bibfnamefont{D.~J.} \bibnamefont{Dean}},
  \bibinfo{author}{\bibfnamefont{C.~W.} \bibnamefont{Johnson}},
  \bibinfo{author}{\bibfnamefont{G.~H.} \bibnamefont{Lang}}, \bibnamefont{and}
  \bibinfo{author}{\bibfnamefont{S.~E.} \bibnamefont{Koonin}},
  \bibinfo{journal}{Phys. Rev. C} \textbf{\bibinfo{volume}{49}},
  \bibinfo{pages}{1422} (\bibinfo{year}{1994}).

\bibitem[{\citenamefont{Rombouts and Heyde}(1998)}]{Rombouts_canonical}
\bibinfo{author}{\bibfnamefont{S.}~\bibnamefont{Rombouts}} \bibnamefont{and}
  \bibinfo{author}{\bibfnamefont{K.}~\bibnamefont{Heyde}}, \bibinfo{journal}{J.
  Comput. Phys.} \textbf{\bibinfo{volume}{140}}, \bibinfo{pages}{453 }
  (\bibinfo{year}{1998}), ISSN \bibinfo{issn}{0021-9991}.

\bibitem[{\citenamefont{Rehman and Ipsen}(2011)}]{Rehman_poly}
\bibinfo{author}{\bibfnamefont{R.}~\bibnamefont{Rehman}} \bibnamefont{and}
  \bibinfo{author}{\bibfnamefont{I.~C.~F.} \bibnamefont{Ipsen}}
  (\bibinfo{year}{2011}), \eprint{arXiv:1104.3769}.

\bibitem[{\citenamefont{Wang et~al.}(2017)\citenamefont{Wang, Assaad, and
  Parisen~Toldin}}]{Assaad2017_canonical}
\bibinfo{author}{\bibfnamefont{Z.}~\bibnamefont{Wang}},
  \bibinfo{author}{\bibfnamefont{F.~F.} \bibnamefont{Assaad}},
  \bibnamefont{and}
  \bibinfo{author}{\bibfnamefont{F.}~\bibnamefont{Parisen~Toldin}},
  \bibinfo{journal}{Phys. Rev. E} \textbf{\bibinfo{volume}{96}},
  \bibinfo{pages}{042131} (\bibinfo{year}{2017}).

\bibitem[{\citenamefont{{Shill, C.R.} and {Drut,
  J.E.}}(2018)}]{Drut2018_canonical}
\bibinfo{author}{\bibnamefont{{Shill, C.R.}}} \bibnamefont{and}
  \bibinfo{author}{\bibnamefont{{Drut, J.E.}}}, \bibinfo{journal}{EPJ Web
  Conf.} \textbf{\bibinfo{volume}{175}}, \bibinfo{pages}{03003}
  (\bibinfo{year}{2018}).

\bibitem[{\citenamefont{Gilbreth and Alhassid}(2015)}]{Gilbreth2015_CPC}
\bibinfo{author}{\bibfnamefont{C.}~\bibnamefont{Gilbreth}} \bibnamefont{and}
  \bibinfo{author}{\bibfnamefont{Y.}~\bibnamefont{Alhassid}},
  \bibinfo{journal}{Comput. Phys. Commun.} \textbf{\bibinfo{volume}{188}},
  \bibinfo{pages}{1 } (\bibinfo{year}{2015}), ISSN \bibinfo{issn}{0010-4655}.

\bibitem[{\citenamefont{Loh~Jr and Gubernatis}(1992)}]{LohJr1992}
\bibinfo{author}{\bibfnamefont{E.~Y.} \bibnamefont{Loh~Jr}} \bibnamefont{and}
  \bibinfo{author}{\bibfnamefont{J.~E.} \bibnamefont{Gubernatis}}, in
  \emph{\bibinfo{booktitle}{Electronic phase transitions (Modern Problems in
  Condensed Matter Sciences)}}, edited by
  \bibinfo{editor}{\bibfnamefont{W.}~\bibnamefont{Hanke}} \bibnamefont{and}
  \bibinfo{editor}{\bibfnamefont{Y.}~\bibnamefont{Kopaev}}
  (\bibinfo{publisher}{North-Holland}, \bibinfo{year}{1992}), pp.
  \bibinfo{pages}{177--235}.

\bibitem[{\citenamefont{Bai et~al.}(2011)\citenamefont{Bai, Lee, Li, and
  Xu}}]{Bai2011_laa}
\bibinfo{author}{\bibfnamefont{Z.}~\bibnamefont{Bai}},
  \bibinfo{author}{\bibfnamefont{C.}~\bibnamefont{Lee}},
  \bibinfo{author}{\bibfnamefont{R.-C.} \bibnamefont{Li}}, \bibnamefont{and}
  \bibinfo{author}{\bibfnamefont{S.}~\bibnamefont{Xu}},
  \bibinfo{journal}{Linear Algebra Appl.} \textbf{\bibinfo{volume}{435}},
  \bibinfo{pages}{659 } (\bibinfo{year}{2011}), ISSN \bibinfo{issn}{0024-3795}.

\bibitem[{\citenamefont{Zwerger}(2012)}]{Zwerger2012}
\bibinfo{editor}{\bibfnamefont{W.}~\bibnamefont{Zwerger}}, ed.,
  \emph{\bibinfo{title}{The {BCS-BEC} Crossover and the Unitary {Fermi} Gas}},
  Lecture Notes in Physics (\bibinfo{publisher}{Springer},
  \bibinfo{year}{2012}).

\bibitem[{\citenamefont{Randeria and Taylor}(2014)}]{Randeria2014_review}
\bibinfo{author}{\bibfnamefont{M.}~\bibnamefont{Randeria}} \bibnamefont{and}
  \bibinfo{author}{\bibfnamefont{E.}~\bibnamefont{Taylor}},
  \bibinfo{journal}{Annual Review of Condensed Matter Physics}
  \textbf{\bibinfo{volume}{5}}, \bibinfo{pages}{209} (\bibinfo{year}{2014}).

\bibitem[{\citenamefont{He et~al.}(2019)\citenamefont{He, Shi, and
  Zhang}}]{Shiwei2019}
\bibinfo{author}{\bibfnamefont{Y.-Y.} \bibnamefont{He}},
  \bibinfo{author}{\bibfnamefont{H.}~\bibnamefont{Shi}}, \bibnamefont{and}
  \bibinfo{author}{\bibfnamefont{S.}~\bibnamefont{Zhang}},
  \bibinfo{journal}{Phys. Rev. Lett.} \textbf{\bibinfo{volume}{123}},
  \bibinfo{pages}{136402} (\bibinfo{year}{2019}).

\bibitem[{\citenamefont{Fang}(2005)}]{Fang2005_thesis}
\bibinfo{author}{\bibfnamefont{L.}~\bibnamefont{Fang}}, Ph.D. thesis,
  \bibinfo{school}{Yale University} (\bibinfo{year}{2005}).

\bibitem[{\citenamefont{Alhassid et~al.}(2008)\citenamefont{Alhassid, Fang, and
  Nakada}}]{Alhassid2008}
\bibinfo{author}{\bibfnamefont{Y.}~\bibnamefont{Alhassid}},
  \bibinfo{author}{\bibfnamefont{L.}~\bibnamefont{Fang}}, \bibnamefont{and}
  \bibinfo{author}{\bibfnamefont{H.}~\bibnamefont{Nakada}},
  \bibinfo{journal}{Phys. Rev. Lett.} \textbf{\bibinfo{volume}{101}},
  \bibinfo{pages}{082501} (\bibinfo{year}{2008}).

\end{thebibliography}

\end{document}